\documentclass[12pt]{article}
\usepackage[margin=1cm]{geometry}
\usepackage{fullpage}
\usepackage{setspace,color}
\usepackage{amsmath,amsfonts,amssymb,amsthm,graphicx,subfigure}

\newcommand{\be}{\begin{equation}} \newcommand{\ee}{\end{equation}}
\newcommand{\bea}{\begin{eqnarray}} \newcommand{\eea}{\end{eqnarray}}
 \def\ba{\begin{array}} \def\ea{\end{array}}

  \def\l{\lambda}

\makeatletter \@addtoreset{equation}{section} \makeatother

\begin{document}

\begin{titlepage}

\vskip -.8cm

\rightline{\small{\tt MCTP-08-57}}

\vskip 1.7 cm

\centerline{\bf \Large Holographic Geometric Entropy at Finite Temperature }

\vskip .4cm
\centerline{\bf \Large from Black Holes in Global Anti de Sitter  Spaces }

\vskip .7 cm


\vskip .2cm \vskip 1cm

\centerline{\large Ibrahima Bah$^1$,  Leopoldo A. Pando Zayas$^{1}$ }

\vskip .5cm

 \centerline{\large and C\'esar A. Terrero-Escalante$^{2}$ }

\vskip 1cm

\vskip .5cm
\centerline{\it ${}^1$ Michigan Center for Theoretical
Physics}
\centerline{ \it Randall Laboratory of Physics, The University of
Michigan}
\centerline{\it Ann Arbor, MI 48109-1040}

\vspace{1cm}
\centerline{\it ${}^2$ Departamento de F\'{\i}sica, Centro de Investigaci\'on y Estudios Avanzados del IPN,}
\centerline{ \it Apdo. Postal 14-740, 07000 M\'exico D.F., M\'exico}

\vspace{1cm}

\begin{abstract}
Using a holographic proposal for the geometric entropy we study its behavior in the geometry of Schwarzschild black holes in global $AdS_p$ for $p=3,4,5$. Holographically, the entropy is determined by a minimal surface. On the gravity side, due to the presence of a horizon on the background, generically there are two solutions to the surfaces determining the entanglement entropy.
In the case of $AdS_3$, the calculation reproduces precisely the geometric entropy of an interval of length $l$ in a two dimensional conformal field theory with periodic boundary conditions. We demonstrate that in the cases of $AdS_{4}$  and $AdS_{5}$ the sign of the difference of the geometric entropies changes, signaling a transition. Euclideanization implies that various embedding of the holographic surface are possible. We study some of them and find that the transitions are ubiquitous.  In particular, our analysis renders a very intricate phase space, showing, for some ranges of the temperature, up to three branches. We observe a remarkable universality in the type of results we obtain from $AdS_4$ and $AdS_5$.
\end{abstract}



\end{titlepage}

\section{Introduction and outlook}

A very natural construction in field theory is the geometric entropy. Given a system in a state $|\Psi>$, one defines a quantity associated with that pure state and a geometrical region $A$ by forming the pure state density matrix $\rho = |\Psi><\Psi|$, tracing over the field variables outside the region to create an impure density matrix $\rho_A = {\rm Tr}_B \rho$, where $B$ is the complement of $A$. One then evaluates the von Neumann entropy $S_A = -{\rm Tr}_A \rho_A \log \rho_A$ associated with this reduced density matrix.

One of the first calculations of this quantity was performed in the eighties \cite{Bombelli:1986rw} for the case of a scalar field propagating in a black-hole background. This paper noticed its proportionality to the area of the black hole horizon. A similar problem but in a strictly field theory sense was posed by Srednicki. Namely, take a ground-state density matrix of a free massless field and trace over the degrees of freedom residing outside an imaginary sphere. The fact that the associated entropy is proportional to the area was connected to the physics of black holes \cite{Srednicki:1993im} and considered as a concrete computation supporting the idea that the entropy area law is a more general phenomenon that originally anticipated.

A lot of activity during the nineties was focused on precise computations of the geometric entropy in various situations \cite{{'tHooft:1984re,Kabat:1994vj,Kabat:1995eq}}. The fact that the geometric entropy came out to be proportional to the area elicited a lot of speculations about its role in the black hole entropy. Perhaps most of that activity was crystalized in a claim by Callan and Wilczek attempting to clarify the relationship between the geometric entropy as defined above and the Bekenstein-Hawking entropy of a black hole. Namely \cite{Callan:1994py}, that geometric entropy is the first quantum correction to a thermodynamic entropy which reduces to the Bekenstein-Hawking entropy in the black hole context. More precisely, the Bekenstein-Hawking entropy and the geometrical entropy are the classical and first quantum contributions to a unified object measuring the response of the Euclidean path integral to the introduction of a conical singularity in the underlying geometry. Some concrete computations followed to support this claim, including the very similar statement: The appropriately defined geometric entropy of a free field in flat space is just the quantum correction to the Bekenstein-Hawking entropy of Rindler space \cite{Holzhey:1994we,Larsen:1994yt,Larsen:1995ax}.

The geometric entropy has turned out to be an important measurement of the entanglement between the two subsystems, it has been extensively studied in this context which includes its use as an order parameter for exotic phases  \cite{exotic}. Recently,  there has been considerable interest in formulating measures of quantum entanglement\footnote{Entanglement entropy and geometric entropy have the same mathematical definition. The main distinction is the situations in which they have been applied. In this note we use entanglement and geometric entropy interchangeably. However, we keep in mind that quantum transitions take place at zero temperature and therefore, in our context, we favor the term geometric entropy.} and applying them to extended quantum systems with many degrees of freedom, such as quantum spin chains.

The AdS/CFT correspondence states roughly that some field theories are mathematically equivalent to string theories containing gravity \cite{Maldacena:1997re}. The question of entanglement was addressed in this context by many authors, including \cite{Brustein:2005vx,Emparan:2006ni}. A crucial breakthrough was taken by Ryu and Takayanagi  \cite{Ryu:2006bv,Ryu:2006ef} who proposed a holographic way of computing it and showed that it precisely matches expectations from two dimensional field theories. Many more works in this area has followed \cite{Barbon:2008ut}-\cite{Fursaev:2006ih}. More recently in \cite{Fujita:2008zv}, a prescription has been suggested for a holographic computation of the geometric entropy. The prescription for the holographic computation of $S_A$ consists on evaluating the area of a minimal surface, in the dual supergravity background, whose boundary is the subsystem $A$:
\be
S_A=\frac{1}{4G_N^{(10)}}\int d^8\sigma e^{-2\phi}\sqrt{G_{ind}^{(8)}}.
\ee
The entropy is obtained by minimizing the above action over all surfaces that approach the boundary of the subsystem $A$.  This is a paradigm shift for no longer the geometric entropy is treated as correction to the black hole entropy as intuitively done in various papers from eighties and early nineties; the statement now is that the geometric field theory entropy can be calculated as a classical quantity in a supergravity background.

\begin{figure}[!ht]
\centering
\includegraphics[width=0.35\linewidth]{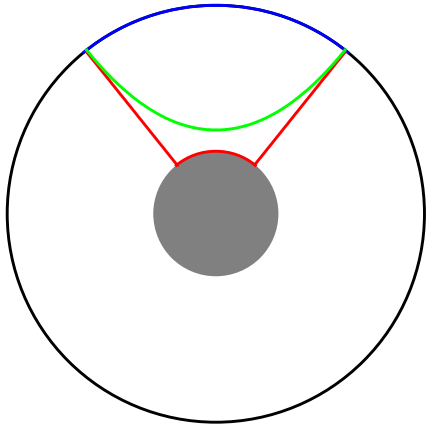}\\
\caption{Two competing configurations for the entanglement entropy in the presence of a black hole horizon. The green
surface represents a continuous configuration while the red surface is piece wise smooth and starts by going straight down from the
boundary to the horizon.  The subsystem $A$ is given by the blue section.  Its characteristic length is $l$.}\label{fig:discon}
\end{figure}

Of particular interest is the case when the supergravity background has a horizon itself. Various groups considered this situation \cite{Brustein:2005vx}, \cite{Emparan:2006ni}. Recently, we revisited the question of entanglement entropy in the presence of horizons
 in which cases there are two surfaces that minimized the above action and therefore a competition ensues among them that we interpret as phase transitions in the field theory (see figure \ref{fig:discon}).  
One of the
surfaces represents a continuous configuration while the second one is piece--wise smooth and extents from the boundary to the horizon.
In appendix \ref{horizon} we prove that this last surface is also minimal. 

For a typical class of embedding and for many types of solutions we showed in \cite{Faraggi:2007fu} that generically there is no transition. The main result of this paper is that after Euclideanization, if we consider a slightly different class of embeddings, motivated largely by \cite{Fujita:2008zv}, then transitions are rather typical for $AdS_4$ and $AdS_5$.

\subsection{Universality of transitions}
Let us state what we consider to be the main result of our paper. {\it The phase diagram of $AdS_4$ and $AdS_5$ is universal, the differences being in the quantitative values.} We briefly summarize it here.  The  metric of the Schwarzschild black hole in global $AdS_{n+1}$ is given as:

\begin{equation}
ds^2 = f_n(r) d\tau^2 + \frac{dr^2}{f_n(r)} + r^2d\Omega_{n-1}^2
\end{equation} where
\begin{eqnarray}
f_n(r) &=& 1 - \frac{M}{r^{n-2}} + \frac{r^2}{L^2} \\
d\Omega_{n-1}^2 &=& \begin{cases} d\theta^2 + \sin^2(\theta) (d\psi^2 + \sin^2(\psi) d\phi^2) \;\;\; \mbox{for} \;\;\; n= 4 \\ d\theta^2 + \sin^2(\theta) d\phi^2 \;\;\; \mbox{for} \;\;\; n= 3 \end{cases}.
\end{eqnarray}

In the case of $AdS_4$, when we fix the $\phi$-cycle we can compute the geometric entropy by embedding the surface as $r(\tau)$ or $r(\theta)$. In the case of  $AdS_5$, we consider $r(\tau)$ and $r(\psi)$.  
\footnote{Later on this paper we discuss this kind of cycles and their relationship with the more typically used temporal cycle.}
Since $\tau$, $\theta$ ($\psi$) have finite period, there exist a maximum $l$ which we denote by  $l_{max}$ and represent with a dashed orange line in the plots below.

\begin{figure}[ht]
\centering
	\subfigure[$y_0>y_0^{crit1}$]{\label{fig:rt1}\includegraphics[scale=.5]{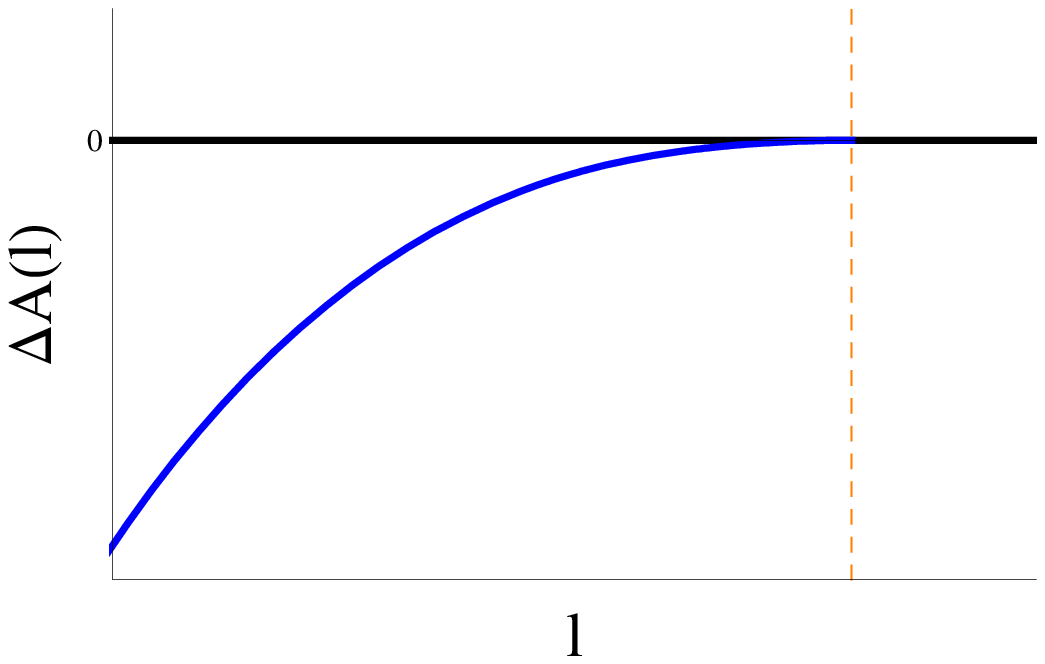}}
	\subfigure[$y_0^{crit1}>y_0>y_0^{crit2}$]{\label{fig:rt2}\includegraphics[scale=.5]{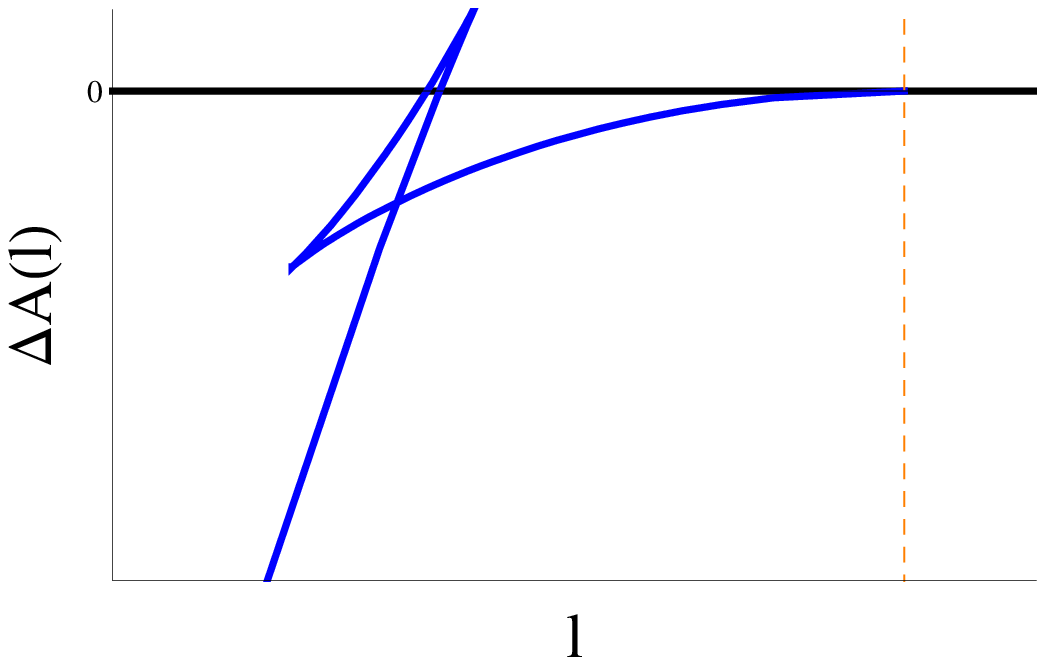}}
	\subfigure[$y_0<y_0^{crit2}$]{\label{fig:rt3}\includegraphics[scale=.5]{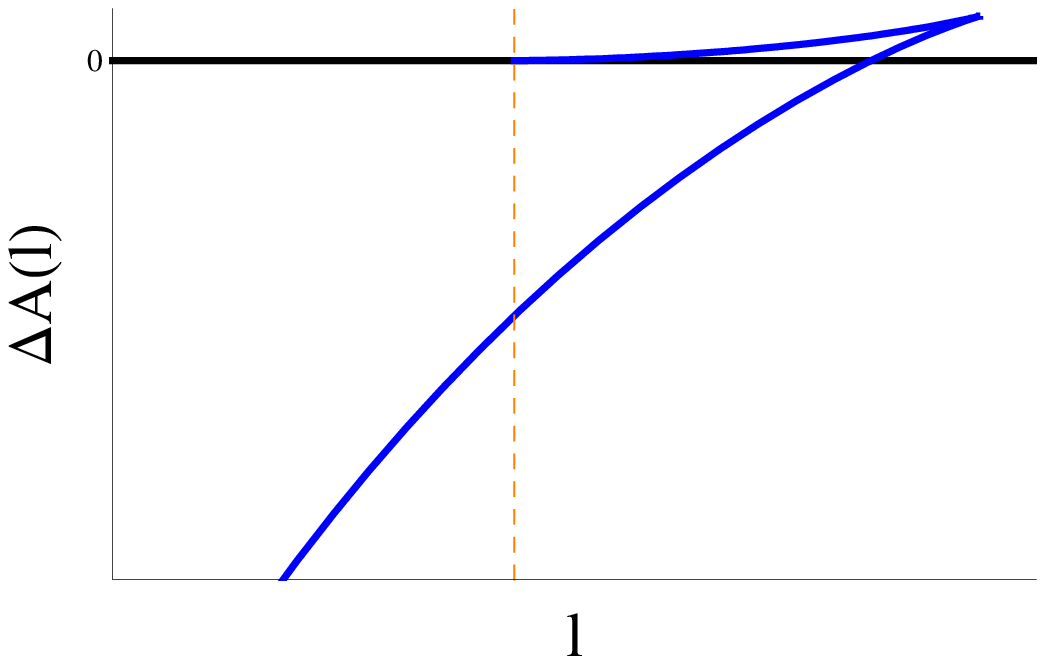}}
 	\caption{For the embedding $r(\tau)$, we plot $\Delta A$ against $l$ at different ranges of $y_0$.  The dashed orange line is the maximum allowed value for $l$.}\label{rt}
\end{figure}

For the $r(\tau)$ embedding, we observe that there exist two critical values of the radius of the black hole horizon or, more physically, two critical temperatures, for which the behavior of the difference of the areas,
$\Delta A(l)=A_{continuous}-A_{piece-wise}$, 
drastically changes.  We tabulate these different behaviors below and denote $y_0=r_{horizon}/L$ the value of the radius of the horizon in units of the $AdS$ radius.
\begin{enumerate}
	\item $y_0 > y_0^{crit1}$, $\Delta A$ has one branch.  In this case, it is smooth and negative as seen in figure \ref{fig:rt1}.  So we do not observe any phase transition.
	\item $y_0^{crit1}>y_0>y_0^{crit2}$, $\Delta A$ has three branches and can change sign.  In this case, $\Delta A$ has two points where it fails to be differentiable.  The first and the third branches intersect creating a loop as seen in figure \ref{fig:rt2}.  This suggests that $\Delta A$ jumps from branch one to branch 3.  Thus we observe a first order phase transition.  Furthermore, this occurs at a negative value of $\Delta A$,  which suggests that this transition occurs in the phase dominated by the smooth surface configuration.
	\item $y_0<y_0^{crit2}$, $\Delta A$ has two branches, one positive and the other negative as seen in figure \ref{fig:rt3}.  The branch where $\Delta A$ is positive occurs for $l> l_{max}$; it is not physical.  The physical branch however, does not go to zero as $l$ approaches $l_{max}$.  This suggest that $\Delta A$ discontinuously jumps to 0 at $l=l_{max}$.
\end{enumerate}

This phase structure is observed in various contexts and seems to be very universal. For example, it is found in the thermodynamics of charged $AdS$ black holes \cite{Chamblin:1999tk}, in the study of Nambu-Goto strings in gravitational backgrounds dual to field theory with flavors \cite{Bigazzi:2008gd}. More importantly, this is the typical phase diagram of a  van Der Waals gas\footnote{We thank C. N\'u\~nez from highlighting this universality to us. We hope to elaborate on this apparent universality in a joint collaboration.}.

\begin{figure}[ht]
\centering
\includegraphics[scale=.6]{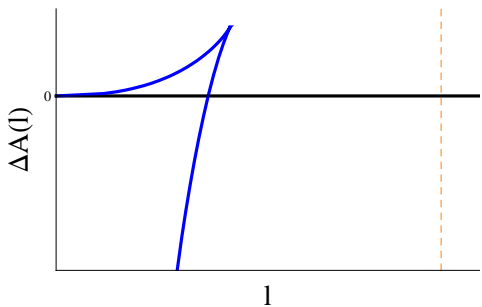}
\caption{$\Delta A$ is plotted against $l$ for the embedding where the radius depends on an angular direction.  We generally observe transition between the smooth and piece-wise smooth surface.  The dashed orange line is the maximum allowed value for $l$.}\label{rtheta}
\end{figure}

When we consider the $r(\theta)$ ($r(\psi)$) embedding, we observe a generic transition between the smooth and piece-wise smooth surfaces, since $\Delta A$ always changes sign at a value of $l$ less than $l_{max}$; see figure \ref{rtheta}.

\subsection{Organization of the paper}

The paper is organized as follows. In section \ref{ads3}, we discussed the geometric entropy in $AdS_3$ and find perfect agreement with the field theory calculation. Essentially, we holographically compute the geometric entropy of a segment of length $l$ in a two dimensional conformal field theory with periodic boundary condition on a circle of radius $l_m$. Sections \ref{ads4} and \ref{ads5} discuss the holographic geometric entropy for the Schwarzschild black hole in global $AdS$ coordinates. There are various possibilities for the embedding of the minimal surface depending on the finite temperature boundary conditions and the form of the dependence of the radial coordinate; we explore a few of them.  In section \ref{sec:hp} we present some comments about the implications of choosing different cycles for the surfaces. We conclude in section \ref{conclusions}.

\section{Geometric entropy in AdS$_3$} \label{ads3}
One of the most interesting results of \cite{Ryu:2006bv} is the holographic calculation of the entanglement entropy for two dimensional conformal field theories. This interesting result has been expanded to a fairly elaborate choice of the subsystem $A$, namely, a discontinuous set of intervals in \cite{Hubeny:2007re}. The starting geometry is the BTZ black hole \cite{Banados:1992wn}:
\begin{equation}
ds^2 = f(r) d\tau^2 + \frac{1}{f(r)} dr^2 + r^2 d\phi^2
\end{equation} where
\begin{equation}
f(r) = 1 - m + \frac{r^2}{L^2} = 1 - m + y^2,
\end{equation} with $y=r/L$.
In the Euclidean version of the metric we have the option of considering $\phi$ as the Euclidean temporal direction.
The subsystem $A$ is then defined as the segment $0\leq \tau \leq \beta l$, where $\beta$ is the period of the $\tau$ cycle and $l \leq 1$. We consider minimal surfaces whose embedding is defined as  $r(\tau)$ with boundary conditions $r(\tau \to 0)=r(\tau \to \beta l) \to \infty$. The metric of the induced surface is then,
\begin{equation}
ds^2 = f(r) (1 + \frac{r'^2}{f^2(r)}) d\tau^2
\end{equation}
with area
\begin{equation}
A_c = \int_{0}^{\beta l} \sqrt{f(r)} \sqrt{1 + \frac{r'^2}{f^2(r)}} d\tau.
\end{equation}
The surface is minimal if $r(\tau)$ satisfies
\begin{equation}
\frac{r'^2}{f^2(r)} = \frac{f(r)}{f(r_*)} -1.
\end{equation}
In the above expression $r_*$ is the turning point of the smooth surface corresponding to $r'(\tau)=0$.
The area can be conveniently rewritten as an integral over $y$
\begin{eqnarray}
A_c&=& \int_{0}^{\beta l} \frac{f(r)}{\sqrt{f(r_*)}} d\tau = 2 L \int_{y_*}^{y_\infty} \frac{1}{\sqrt{f(y) -f(y_*)}} dy.
\end{eqnarray}
For the manifolds we are considering, there is another solution to the boundary problem which consists of a piece-wise smooth surface. The
 area of such surfaces is given by
\begin{equation}
A_{p}= 2 \int_{r_0}^{r_\infty} \frac{dr}{\sqrt{f(r)}}.
\end{equation}
The difference in area is then
\begin{equation}
\Delta A =A_c-A_p= 2 L \left[\int_{y_*}^{y_\infty} \left(\frac{1}{\sqrt{f(y) -f(y_*)}}- \frac{1}{\sqrt{f(y)}} \right) dy - \int_{y_0}^{y_*} \frac{dy}{\sqrt{f(y)}} \right].
\end{equation}
The separation at the boundary $l$ depends on the turning point of the surface in the bulk $y_*$ as follows:
\begin{equation}
l = \frac{2 L}{\beta} \int_{y_*}^{y_\infty} \frac{\sqrt{f(y_*)}}{f(y)\sqrt{f(y) - f(y_*)}} dy.
\end{equation}
These expressions can be integrated analytically to yield
\begin{eqnarray}
\Delta A = 2 L \ln \left(\frac{y_0}{y_*}\right), \qquad
l(y_*) = \frac{1}{\pi}\arcsin\left(\frac{y_0}{y_*}\right), \label{ads3l_ys}
\end{eqnarray}from which we obtain
\begin{equation}
\Delta A(l) = 2L \ln (\sin(\pi l)).
\end{equation}
Note that $\Delta A \leq 0$. Therefore, we observe the entropy must be given by the smooth surface in AdS$_3$.

We can rewrite the expression for the entropy in terms of intrinsically field theoretic quantities. Recall that the central charge in the two dimensional CFT can be related to Newton's constant $G$ and the radius of $AdS_3$ (which is denoted by $L$) as:
$c =  \frac{3 L}{2 G}$.

Since the holographic entanglement entropy is given as $S = \frac{A}{4G}$, we obtain
\begin{equation}
\Delta S = \frac{\Delta A}{4G} = \frac{L}{2G}\ln (\sin(\pi l)) = \frac{c}{3} \ln (\sin(\pi l)).
\end{equation}

Interestingly, equation (\ref{ads3l_ys}) shows that the length $l$ is bounded by $1/2$ and at that value the difference in entropies vanishes. The difference of entropies is symmetric under  $l \to 1-l$, this can be interpreted as meaning that the entropy of the subsystem $A$ defined by the segment $l$ and its complement, defined by the segment $1-l$ are the same. The equality $S_A=S_B$, for the entropy of a subsystem $A$ and its complement $B$, is a general property of the entanglement entropy and it is true in higher dimension. It is also worth noting that the difference in entropy is independent of $y_0$, making it independent of the periodicity along the original temporal direction $\tau$.  This property changes when we go to higher dimensions.  In fact, it will allow for an interesting phase structure.

Our holographic result perfectly agrees with the entanglement entropy of a subsystem of length $l$ in a finite system with periodic boundary condition \cite{Holzhey:1994we,Calabrese:2004eu} and finite length. The precise field theoretic expression is
\begin{equation}
\label{2dentropycompact}
S(l) = \frac{c}{3} \ln\left(\frac{l_m}{\pi a}\sin (\frac{\pi l}{l_{m}})\right) + c_1
\end{equation}
where $l_{m}$ is the size of the system, $c_1$ is a non-universal constant and $a$ is the lattice spacing corresponding to a UV cut-off.

Our result from the continuous surface is
\begin{equation}
A_c=2L \ln \left(\frac{y_\infty}{y_0}\sin\pi l\right) \longrightarrow
S(l)=\frac{c}{3} \ln\left(\frac{\beta y_\infty}{2\pi L} \sin (\pi l)\right) = \frac{c}{3} \ln\left(\frac{ y_\infty}{\pi } \sin (\pi l)\right) + \frac{c}{3} \ln\left(\frac{\beta}{2 L}\right).
\end{equation}
This expression captures the field theory answer (\ref{2dentropycompact}) precisely.

\section{Geometric entropy from AdS$_4$} \label{ads4}

In this section we analyze the geometric entropy in the holographic context of the Schwarzschild black hole in $AdS_4$. The dual field theory is expected to be a $2+1$  superconformal field theory at some inverse temperature $\beta$ living on a the conformal boundary of $AdS_4$ which is basically a 2-sphere. Although the precise conformal field theory can be understood in terms of the AdS/CFT correspondence, our main motivation comes from the fact that there seems to be a sort of universality for field theories in $2+1$ dimensions that make some of our results relevant for a wide class of problems. For example, systems whose gravity dual  asymptote to $AdS_4$ have been recently considered in connection with various condensed matter systems \cite{Herzog:2007ij,Hartnoll:2007ai,Hartnoll:2007ih}.

The Euclidean geometry is given by:
\begin{equation}
ds^2 = f(r) d\tau^2 + \frac{1}{f(r)} dr^2  + r^2 (d\theta^2 + \sin^2(\theta) d\phi^2 ), \qquad f(r) = 1 - \frac{M}{r} + \frac{r^2}{L^2}.
\end{equation}
Using the convenient coordinate  $y=r/L$, we have: $f(y) = \frac{1}{y} (y^3 + y -m)$ where $m=M/L$. Defining the position of the horizon as the largest root of $f(y_0)=0$ we find that to avoid conical singularities the dimensionless periodicity in the $\tau$ direction must be
\begin{equation}
\tilde{\beta} = \frac{\beta }{L} = \frac{4 \pi y_0}{1+3y_0^2}.
\end{equation}

In the Euclideanized form of the metric we can clearly choose to embedded the surface at fixed $\tau$ or at fixed $\psi$. We have considered embedding with fixed $\tau$ previously \cite{Faraggi:2007fu}. In this paper we focused on embeddings where another coordinate is kept fixed. Namely, considered embeddings with $\psi$ fixed. Then we have two options for the embedding of the surface, $r(\theta)$ and $r(\tau)$. We proceed to explore both of them in what follows (see Section \ref{sec:hp} for comments on these embeddings).

\subsection{$r(\theta)$ minimal surface}

The induced metric for this embedding is,
\begin{equation}
ds^2 = f(r) d\tau^2 + r^2 (1 + \frac{r'^2}{r^2 f(r)}) d\theta^2,
\end{equation}
where the prime means derivative with respect to $\theta$ and $0\leq \theta \leq \pi l$. The area is given by:
\begin{equation}
A_c = \beta \int_0^{ \pi l} r \sqrt{f(r)} \sqrt{1 + \frac{r'^2}{r^2 f(r)}} d\theta.
\end{equation}
The equation of motion can be written as
\begin{equation}
\frac{r'^2}{r^2 f(r)} = \frac{r^2 f(r)}{r_*^2 f(r_*)} -1.
\end{equation}
From the equation of motion, we can rewrite the area as,
\begin{eqnarray}
A_c &=& \beta \int_0^{ \pi l} \frac{r^2 f(r)}{r_* \sqrt{f(r_*)}} d\theta =
2 \beta \int_{r_*}^{r_\infty} \frac{r \sqrt{f(r)}}{\sqrt{r^2 f(r) -r_*^2 f(r_*)}} dr .\label{areacads4theta}
\end{eqnarray}
The area of the piece-wise smooth surface that also solves the boundary condition problem is
\begin{equation}
A_p = 2 \beta \int_{r_0}^{r_\infty} dr = 2 \beta (r_\infty - r_0).
\end{equation}
The difference in areas, $A_c-A_p$, is
\begin{eqnarray}
\Delta A &=& 2 \beta L \left[ \int_{y_*}^{y_\infty} \left(\frac{y \sqrt{f(y)}}{\sqrt{y^2 f(y) -y_*^2 f(y_*)}} -1\right) dy - (y_* -y_0) \right].
\end{eqnarray}
The relation between the size of subsystem $A$, $l$, and the minimum of the smooth surface $y_*$ is
\begin{equation}
l ( y_*) = \frac{2}{\pi} \int_{y_*}^{y_\infty} \frac{y_* \sqrt{f(y_*)}}{y \sqrt{f(y)} \sqrt{y^2 f(y) -y_*^2 f(y_*)}} dy.\label{lyads4theta}
\end{equation}
Our goal is to show that the difference in entropy changes sign. We managed to do that analytically by studying some limits of the above expression. First, we consider the limit when the smooth surface drops into the bulk almost all the way to the horizon. In this case we can show (see appendix \ref{app:rtheta}) that
\begin{equation}
 \Delta A(y_* = y_0 + \epsilon) > 0
\end{equation}
where $\epsilon << y_0$.
In the other limit, when the surface remains in the large $y$ region, we can simplify the expression using $x=y/y_*$ to the form:
\begin{eqnarray}
\Delta A &=& 2 \beta L y_* \left[ \int_{1}^{x_\infty} \left(\frac{x^2}{\sqrt{x^4 -1}} -1\right) dx - 1 \right] \\
&=& 2 \beta L y_* \left[0.4009 - 1 \right]<0.
\end{eqnarray}

\begin{figure}[ht!]
\centering
	\subfigure[]{\label{fig:ads4ltheta}\includegraphics[scale=.6]{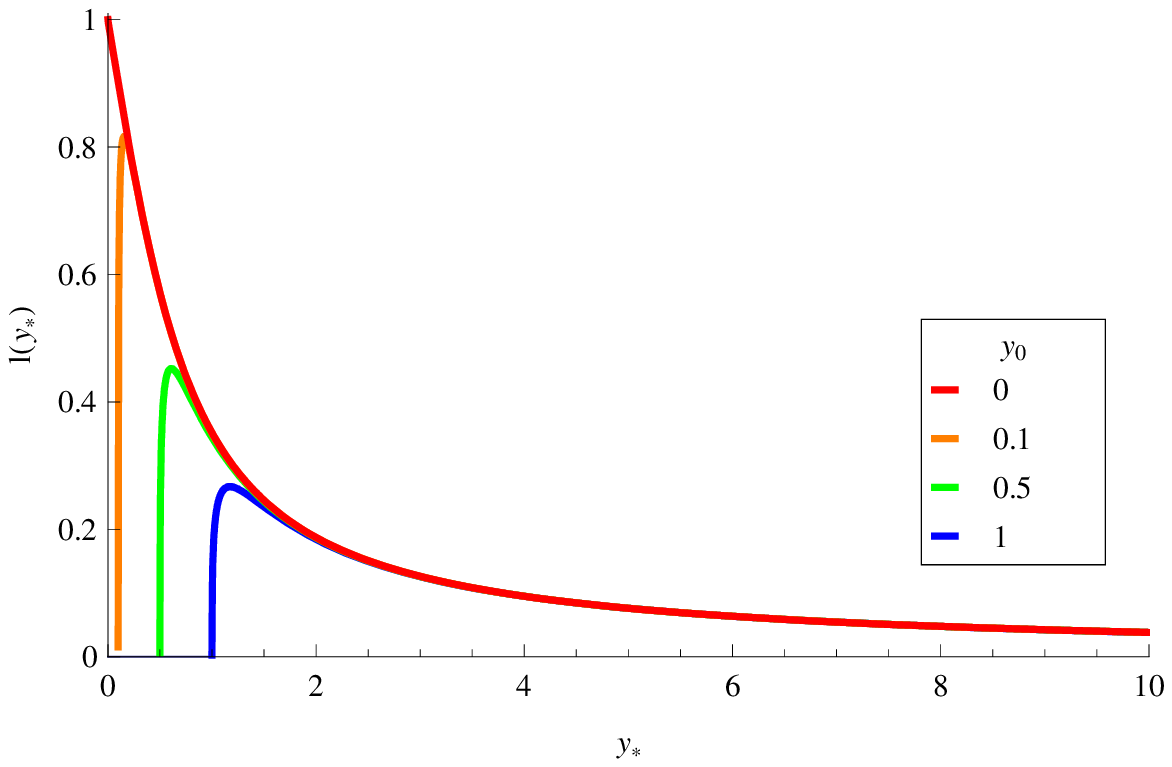}}
	\subfigure[]{\label{fig:ads4atheta}\includegraphics[scale=.6]{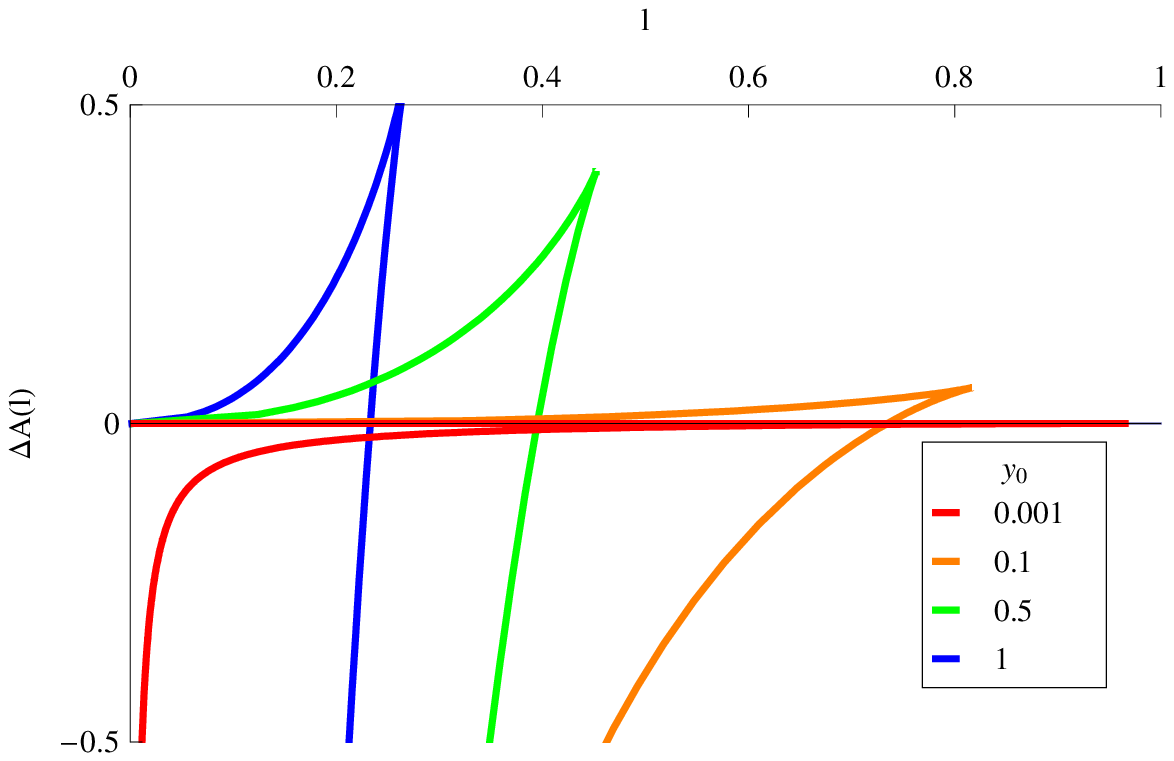}}
 	\caption{\ref{fig:ads4ltheta} shows the function $l(y_*)$ at $y_0 = 0, 0.1, 0.5, 1$.  For the $r(\theta)$ embedding, we observe that $l$ is strictly bounded by 1.  \ref{fig:ads4atheta} shows the behavior of $\Delta A$ as a function of $l$.  We generically observe a transition.}\label{ads4theta}
\end{figure}

We thus observe that the difference in area changes sign.  However, this is not enough to claim transition as the value of $l$ must be bounded in order to have a well defined problem on a circle of finite size.  Given our normalization we can claim the existence of a transition only when there exist a $y_*^c$ such that $\Delta A (y_*^c) =0$ and the length $l$ evaluated at that value of $y_*^c$ satisfies $l(y_*^c)<1$.  We can then supplement the above statement by showing that the function $l(y_*)\leq 1$ for all black hole sizes, that is, for all values of $y_0$.  This is observed in figure \ref{fig:ads4ltheta} where $l$ is plotted against $y_*$ at $y_0 = 0, 0.1, 0.5, 1$.  In fact, it is numerically observed that $1>l(y_*,y_0^a)>l(y_*,y_0^b)$ if $y_0^a<y_0^b$.  When $y_0 =0$, which corresponds to global $AdS$, there  is no conical singularity problem; thus the period of $\tau$ can take on any value.  If the period is not zero, it is clear that $\Delta A$ will change sign since the above argument only requires a non-zero period for $\tau$.

We conclude that the difference in area changes sign for all finite $y_0$. For completeness we also plot, in figure \ref{fig:ads4atheta}, the difference in areas as a function of the separation length $l$, that is,  $\Delta A(l)$ for various values of $y_0$.

Let us finish this subsection with a comment about the ultraviolet behavior of the entropy. In the large $y_*$ limit,  (\ref{areacads4theta}) and (\ref{lyads4theta}) integrate to
\begin{eqnarray}
A_c = 2 \tilde{\beta} L^2 \left( y_\infty - \frac{\sqrt{\pi}\Gamma(\frac{3}{4})}{\Gamma(\frac{1}{4})} y_* \right), \quad
l(y_*) = \frac{2}{\pi} \frac{1}{y_*} \frac{\sqrt{\pi}\Gamma(\frac{3}{4})}{\Gamma(\frac{1}{4})}.
\end{eqnarray}
The expression for the area of the smooth surface in this limit is

\begin{equation}
A_c(l) =  2 \tilde{\beta} L^2 \left( y_\infty - \left[\frac{\sqrt{\pi}\Gamma(\frac{3}{4})}{\Gamma(\frac{1}{4})}\right]^2 \frac{2}{\pi l} \right).
\end{equation}

We can then write the entropy, $S = \frac{A}{4G_N^{(4)}}$, in terms of the Planck constant, $l_p$, and the number of $M2$ branes, $N$, from the relations
\begin{equation}
G_N^{(11)} = 16 \pi^7 l_p^9, \;\;\; G_N^{(4)} = \frac{G_N^{(11)}}{R_{s7}^7 \Omega_7}, \;\;\; 2 L = R_{s7} = l_p {32 \pi N}^{1/6}.
\end{equation} The entropy is then

\begin{equation}
S_c(l) = \frac{4 \Omega_7}{(2\pi)^7} (32 \pi^2 N)^{7/6} \,\,\frac{\beta r_\infty}{l_p^2} - \frac{8 C^2 \Omega_7}{(2\pi)^8} \frac{(32 \pi^2 N)^{4/3}}{l_p} \frac{\beta}{l}
\end{equation} where $C =\left[\frac{\sqrt{\pi}\Gamma(\frac{3}{4})}{\Gamma(\frac{1}{4})}\right] $.  We notice that there is a linear divergent piece, we might identify $r_\infty/l_p^2$ with an ultraviolet cutoff. It is interesting to note that the power of $N^{7/6}$ is different from the naively expected $N^{3/2}$. The simplest reason comes from dimensional analysis, that is, having the new scales $\beta$ and $r_\infty$ in the problem makes us deviate from the conformal result of $N^{3/2}$. Namely, each length scale should have contributed $N^{1/6}$ in the conformal limit, we can simply verify that $N^{3/2}=N^{7/6}N^{2/6}$. We are optimistic that $N^{7/6}$ versus $N^{3/2}$ might have a real physical explanations in terms of degrees of freedom involved in the entanglement. In fact we will see similar deviations from the standard conformal scaling in all the problems that we tackle in this paper.

The entropy from the piece-wise smooth surface is also
\begin{equation}
S_p = \frac{4 \Omega_7}{(2\pi)^7} \frac{(32 \pi^2 N)^{7/6}}{l_p^2} \beta r_\infty - \frac{\Omega_7}{3 (2\pi)^6}(32 \pi ^2 N)^{3/2} \left(1+ \left(1 - \frac{3}{\pi} \frac{\beta^2}{l_p^2 (32 \pi ^2 N)^{1/3}} \right)^{1/2} \right)
\end{equation} where we have used the fact that $r_0 = \frac{4 \pi L^2}{3 \beta}\left(1+ \left(1 - \frac{3 \beta^2}{4 \pi^2 L^2} \right)^{1/2} \right)$.

We observe that the divergent piece is the same in both cases, as expected.  Furthermore this quantity is proportional to the area of the boundary of subsystem $A$, $\beta r_\infty$.  This is in agreement with the area law discussed in \cite{Ryu:2006ef} and \cite{Srednicki:1993im}.  We see that $S_p$ does not depend on $l$. Therefore, when the entropy difference changes sign $\frac{\partial S_A}{\partial l}$ experiences a discontinuity.

\subsection{$r(\tau)$ minimal surface}

The induced metric for this surface is given by
\begin{equation}
ds^2 = f(r) (1 + \frac{r'^2(\tau)}{f^2(r)}) d\tau^2 + r^2 d\theta^2,
\end{equation}
where the prime stands for derivative with respect to $\tau$ and $0 \leq \tau \leq \beta l$. The area of the embedded surface
\begin{equation}
A_c=\pi \int_{0}^{\beta l} r \sqrt{f(r)} \sqrt{1 + \frac{r'^2(\tau)}{f^2(r)}} d\tau.
\end{equation} The equation of motion for the minimal surface is then
\begin{equation}
\frac{r'{}^2(\tau)}{f^2(r)} = \frac{r^2 f(r)}{r_*^2 f(r_*)} -1.
\end{equation}
with area
\begin{eqnarray}
\label{rtac}
A_c &=& \pi \int_{0}^{\beta l} \frac{r^2 f(r)}{r_* \sqrt{f(r_*)}} d\tau = 2 \pi \int_{r_*}^{r_\infty} \frac{r^2 } { \sqrt{r^2 f(r) - r_*^2 f(r_*)}} dr.
\end{eqnarray}
The area of the piece-wise smooth surface is also,
\begin{equation}
A_p = 2\pi \int_{r_0}^{r_\infty} \frac{r dr}{\sqrt{f(r)}} \overbrace{=}^{r_0>>L}  2\pi L r_\infty - C \frac{16 \pi^2 L^3}{\beta}
\end{equation} where $C = \frac{1}{3}\left(1 - \int_{0}{\infty}\frac{x^{3/2} - \sqrt{x^3 -1}}{\sqrt{x^3 -1}} dx \right) \approx 0.144$.
 We then obtain the difference in areas $A_c-A_p$ as
\begin{eqnarray}
\Delta A &=& 2 \pi L^2 \left[ \int_{y_*}^{y_\infty} \left(\frac{y^2 } { \sqrt{y^2 f(y) - y_*^2 f(y_*)}} - \frac{y }{\sqrt{f(y)}}\right) dy - \int_{y_0}^{y_*} \frac{y dy}{\sqrt{f(y)}} \right].
\end{eqnarray}
The length $l$ of the subsystem in this case is
\begin{equation}
\label{rtly_s}
l = \frac{2L}{\beta}  \int_{y_*}^{y_\infty} \frac{y_* \sqrt{f(y_*)}}{f(y) \sqrt{y^2 f(y) - y_*^2 f(y_*)}} dy.
\end{equation}
Let us first consider the limiting regions of the above expressions.
We observe that
\begin{equation}
 \Delta A(y_* = y_0)= 0.
\end{equation}
For large $y_*$, we can write
\begin{eqnarray}
\Delta A &=& 2 \pi L^2 \left[ y_* \int_{1}^{x_\infty} \left(\frac{x^2}{\sqrt{x^4 -1}} -1\right) dx - \int_{y_0}^{y_*} \frac{y dy}{\sqrt{f(y)}}  \right] \\
&=& 2 \pi L^2 y_* \left[0.4009 - 1 \right]<0.
\end{eqnarray}  In the last line, we used the fact that the second integral is $y_*$ in the large $y_*$ limit.

\begin{figure}[ht!]
\centering
		\subfigure[]{\label{fig:ads4tl-a}\includegraphics[scale=.65]{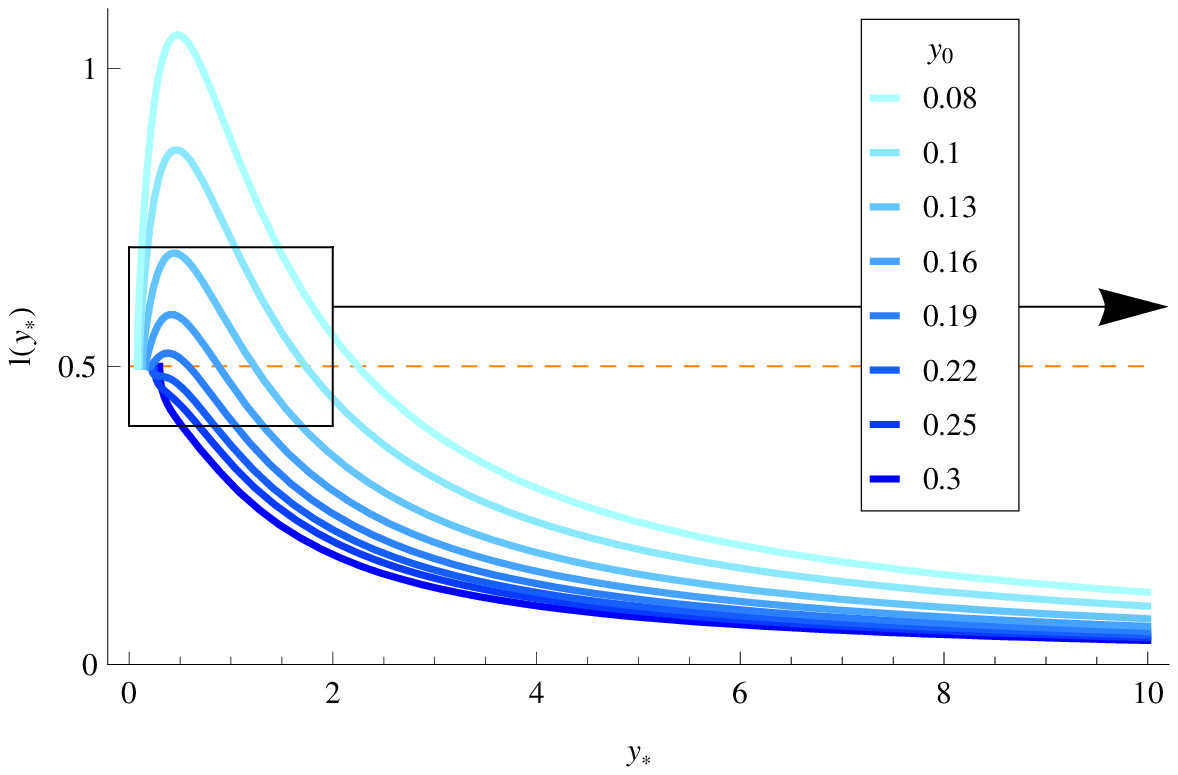}}
		\subfigure[]{\label{fig:ads4tl-b}\includegraphics[scale=.65]{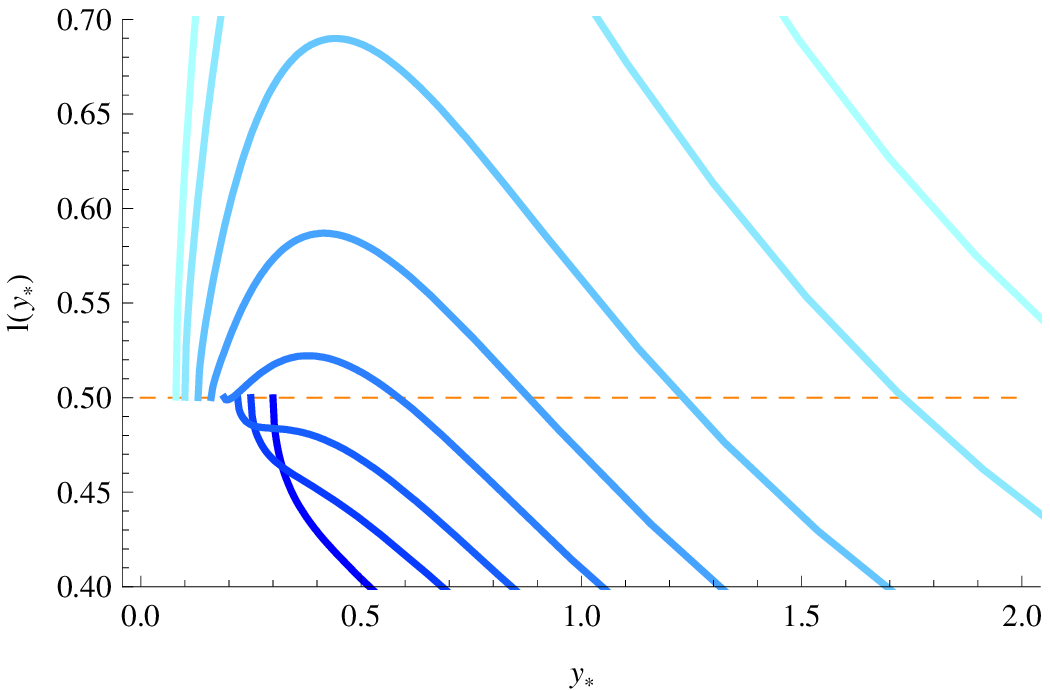}}
 	\caption{In \ref{fig:ads4tl-a} we have plotted $l$ against $y_*$ at various values of $y_0$ and in \ref{fig:ads4tl-b}, we have zoomed in the region around $l=.5$. }
	\label{fig:ads4tl}
\end{figure}

\begin{figure}[h!]
\centering
		\subfigure[]{\label{fig:ads4ta-a}\includegraphics[scale=.60]{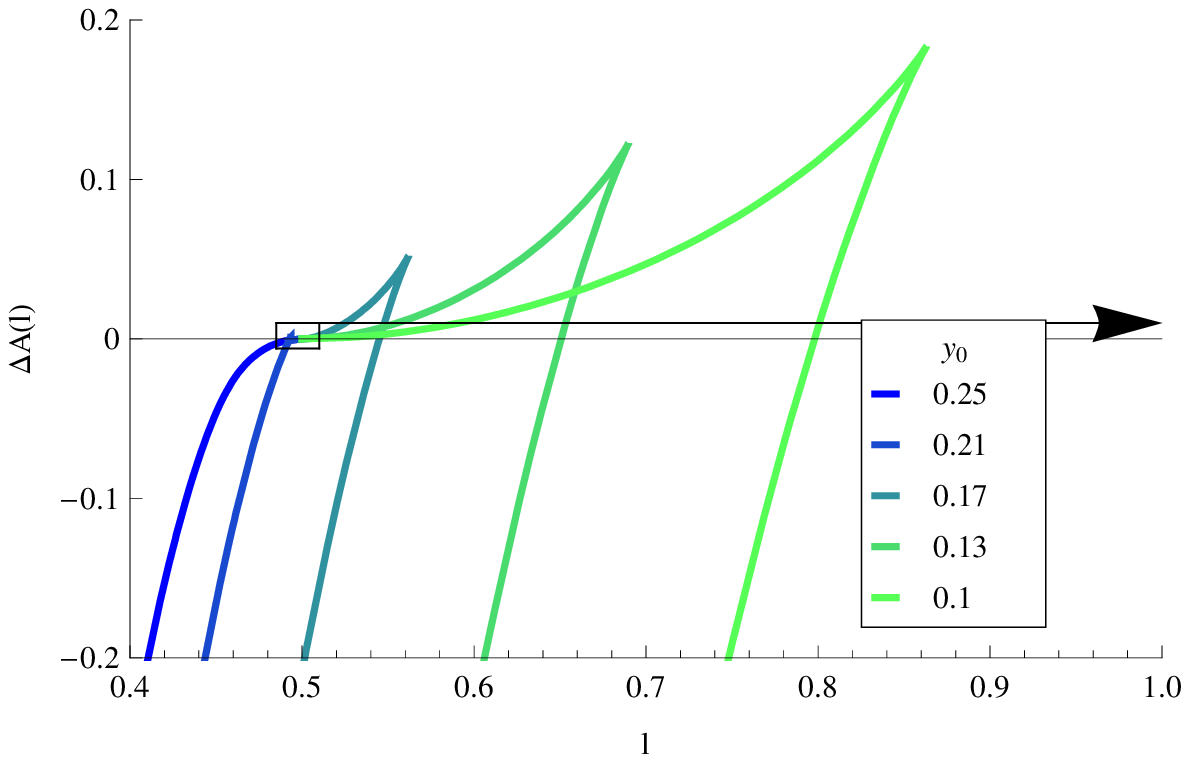}}	 	  	
		\subfigure[]{\label{fig:ads4ta-b}\includegraphics[scale=.60]{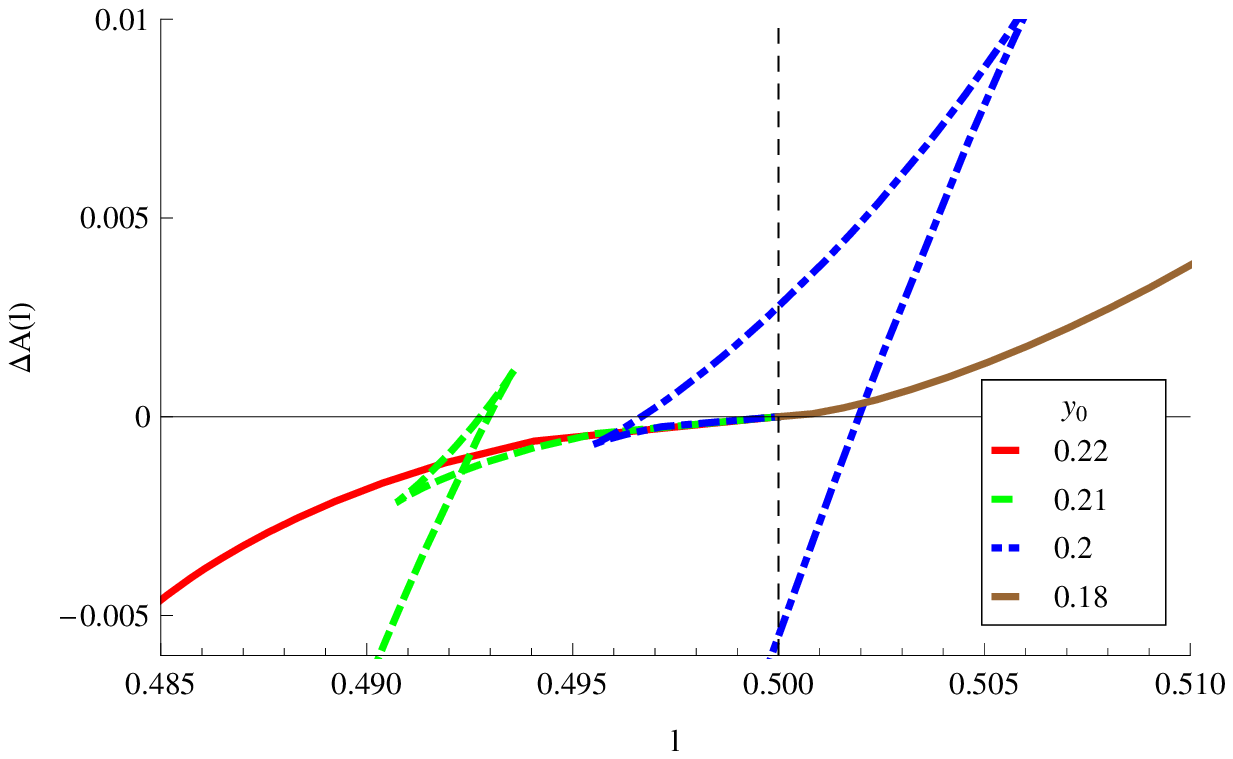}}
		\subfigure[]{\label{fig:ads4ta-c}\includegraphics[scale=.65]{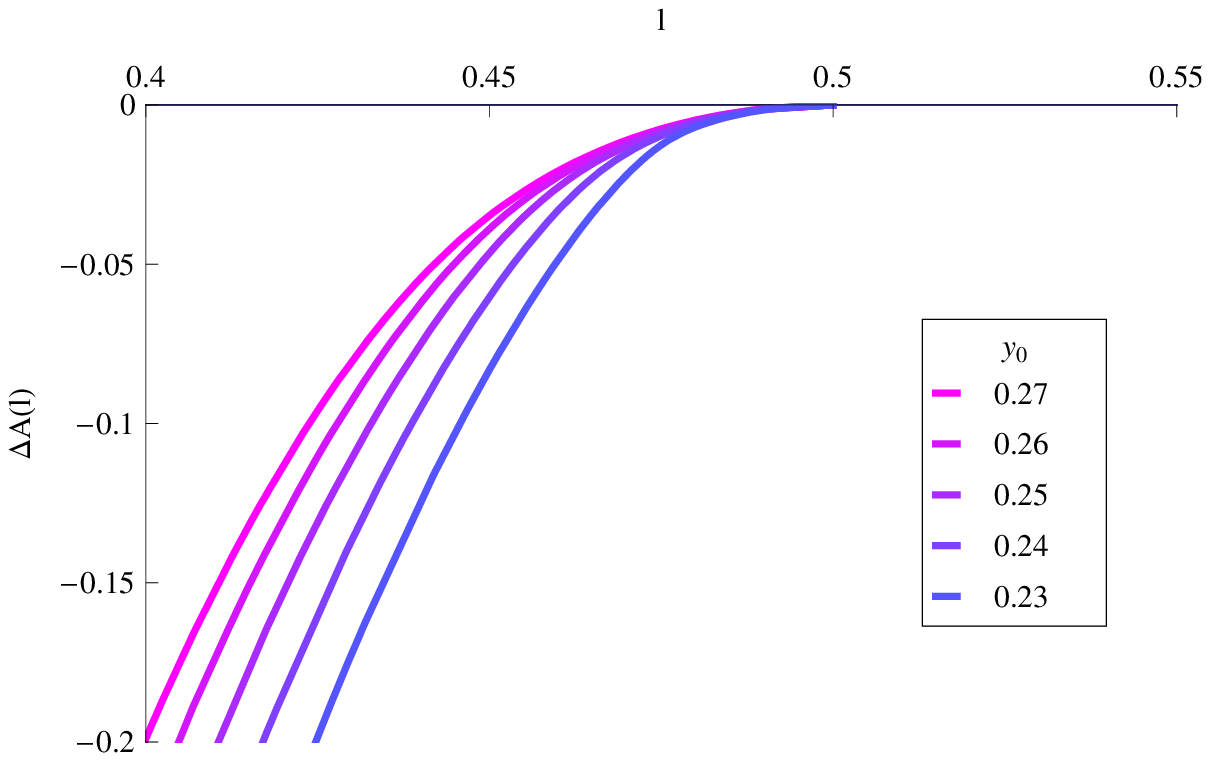}}
	\caption{In \ref{fig:ads4ta-a}, we plot the difference in area for values of $y_0$ less that .189.  In \ref{fig:ads4ta-b}, we have zoomed in the region around $l=.5$ to show interesting transition structure.  In \ref{fig:ads4ta-b}, we plot the difference in area for values of $y_0$ greater than .217.}
	\label{fig:ads4ta}
\end{figure}

\begin{figure}[ht!]
\centering
		\subfigure[$y_0 = .22$]{\label{fig:ads4tla-a}\includegraphics[scale=.5]{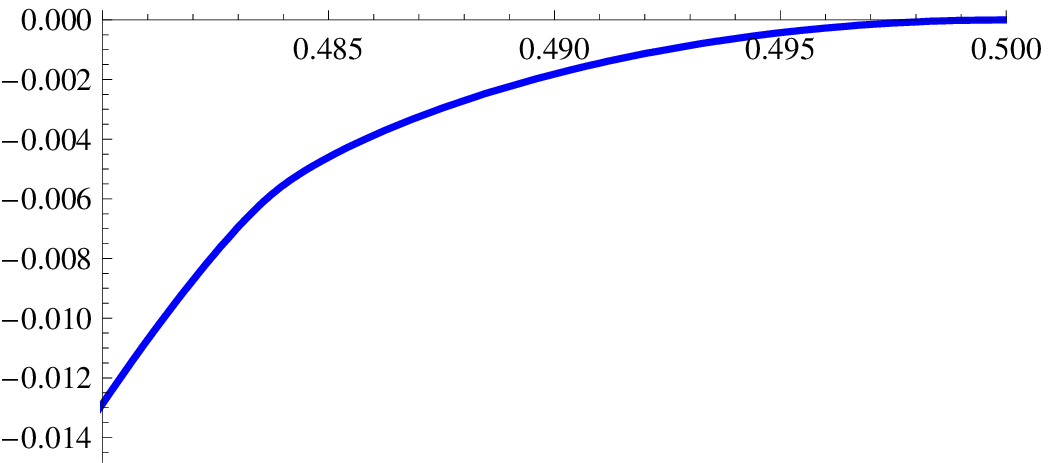}}
		\subfigure[$y_0 = .217$]{\label{fig:ads4tla-b}\includegraphics[scale=.5]{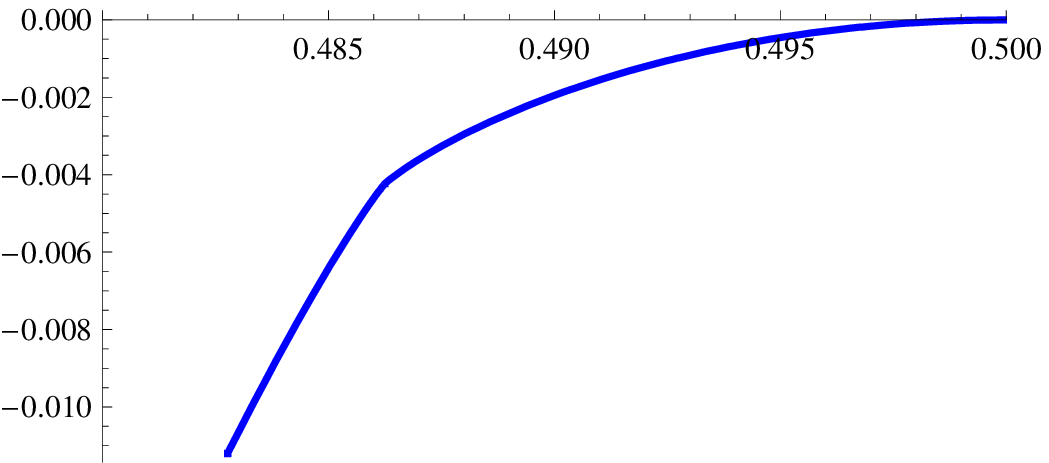}}	 	  	
		\subfigure[$y_0 = .214$]{\label{fig:ads4tla-c}\includegraphics[scale=.5]{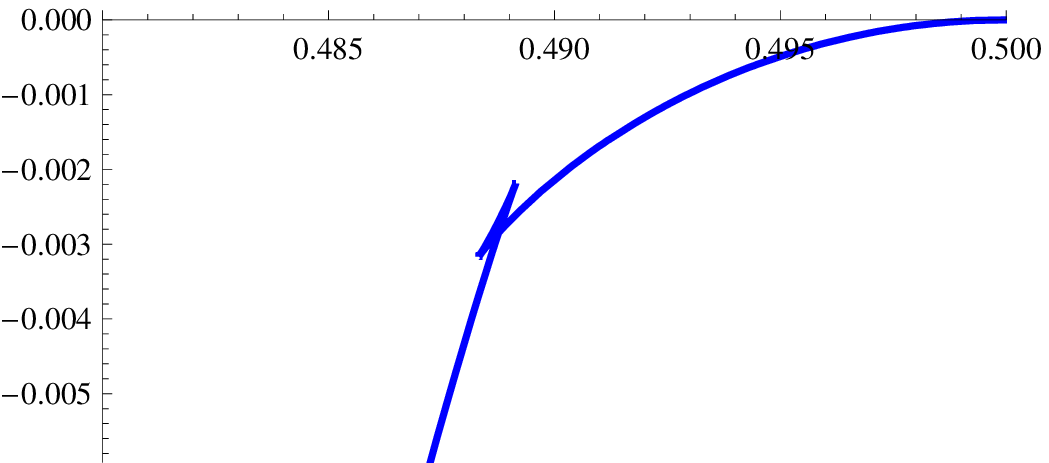}}
		\subfigure[$y_0 = .212$]{\label{fig:ads4tla-d}\includegraphics[scale=.5]{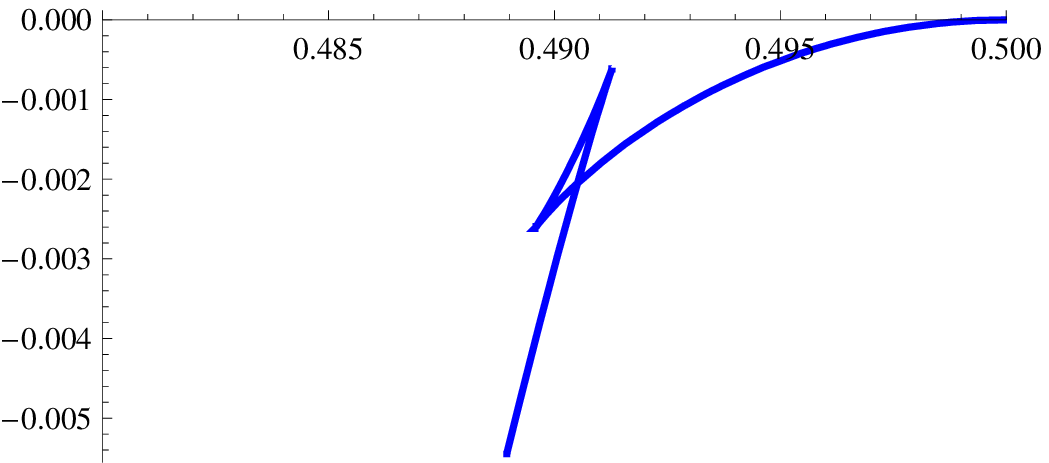}}	
		\subfigure[$y_0 = .21$]{\label{fig:ads4tla-e}\includegraphics[scale=.5]{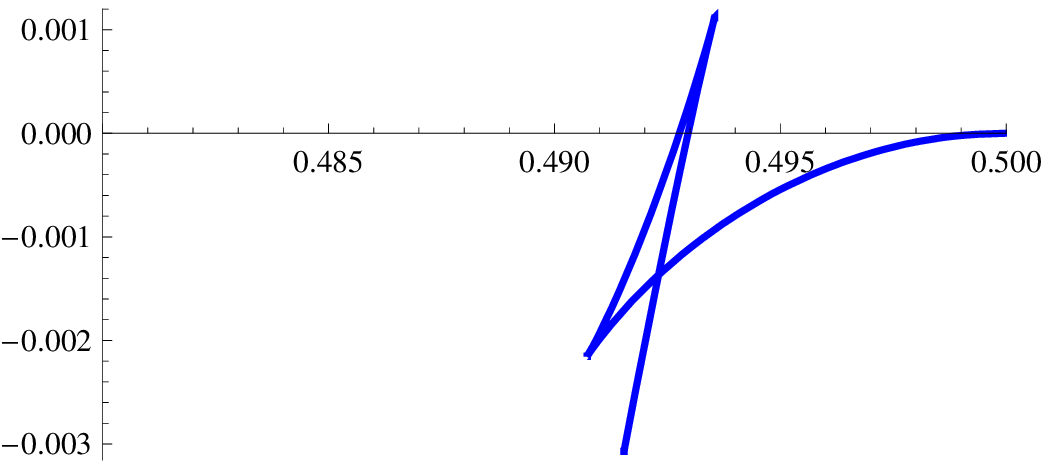}}	
		\subfigure[$y_0 = .208$]{\label{fig:ads4tla-f}\includegraphics[scale=.5]{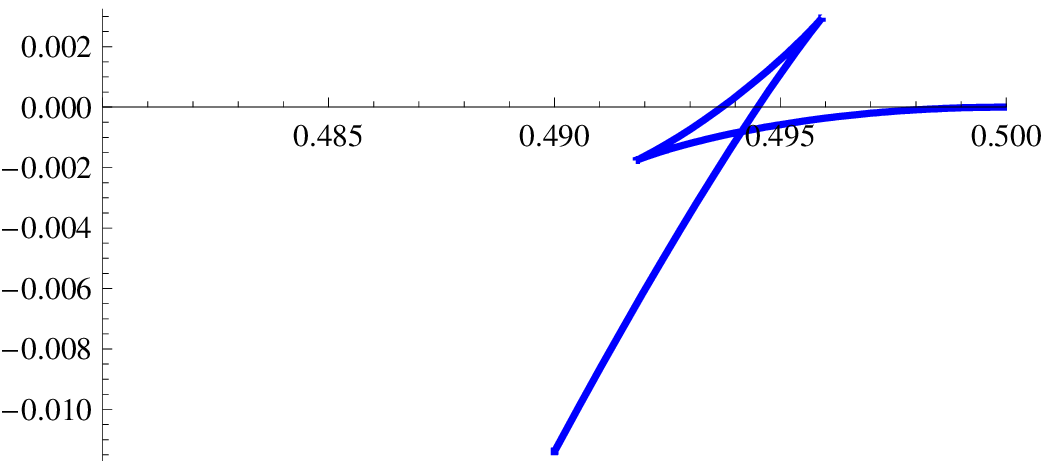}}	
		\subfigure[$y_0 = .202$]{\label{fig:ads4tla-g}\includegraphics[scale=.5]{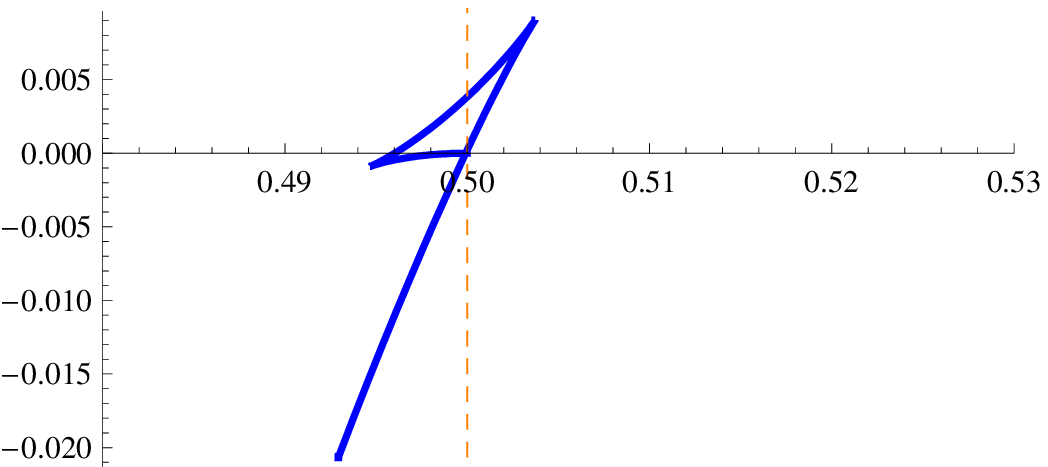}}	
		\subfigure[$y_0 = .2$]{\label{fig:ads4tla-h}\includegraphics[scale=.5]{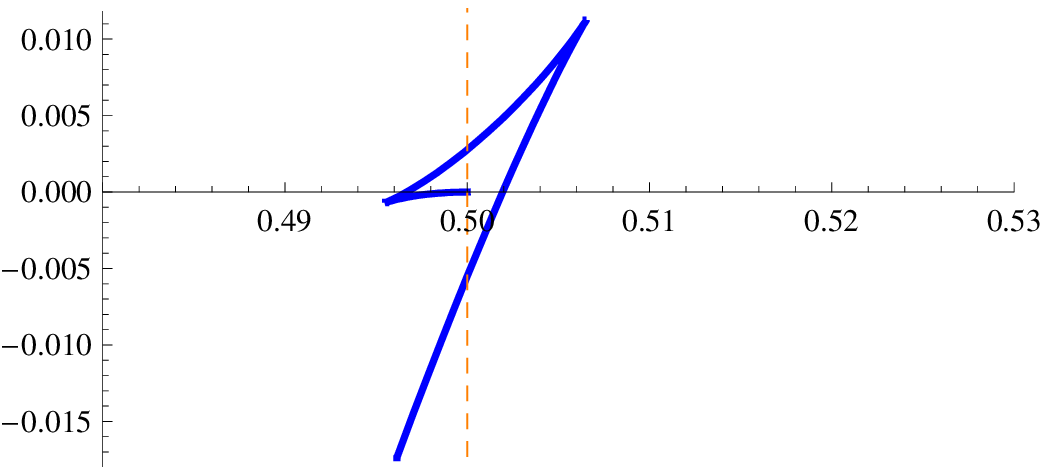}}	
		\subfigure[$y_0 = .186$]{\label{fig:ads4tla-i}\includegraphics[scale=.5]{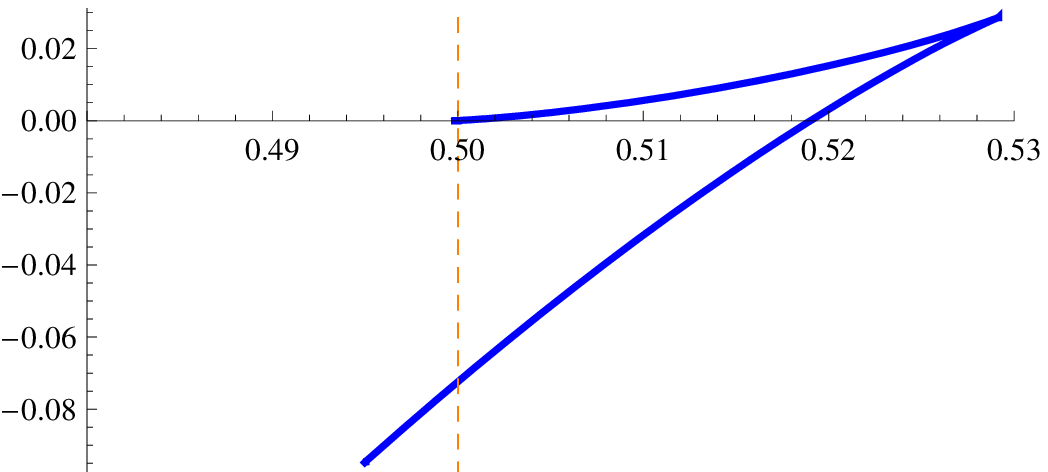}}	
												
	\caption{We plot the difference in area for $r(\tau)$ in $AdS_4$ vs. $l$ at difference values of $y_0$.  These snapshots track the the difference in entropy as it experiences a first order phase transition.}
	\label{fig:ads4tla}
\end{figure}

The analysis above does not show whether the difference in entropy changes sign nor does it show whether there exist a transition.  In figure \ref{fig:ads4tl} and \ref{fig:ads4ta}, we have plotted $l$ vs. $y_*$ and $\Delta A$  vs. $l$ at appropriate values of $y_0$.  We generically observe that $l(y_0)=\frac{1}{2}$ from the numerical analysis. In appendix \ref{sec:limitlaty0} we proved analytically that as $y_*\to y_0$ the length $l\to 1/2$.  This feature is consistent with the fact that $y_0$ is really the origin of this background and since it is non thermal, the difference in area must satisfy $\Delta A (l) \equiv \Delta A (1-l)$.  Thus we should not consider cases for $l>0.5$.  This was also observed for AdS$_3$.
Furthermore, numerical analysis show interesting behavior for $\Delta A(l)$.  We observe that $\Delta A$ is negative and monotonic for $y_0 > 0.217$ (see figure \ref{fig:ads4ta-c}),  then from $y_0 = 0.217$ to $y_0=0.189$, $\Delta A$ and $l$ both have three branches.  Within this interval, $\Delta A$ can change sign.  Then from $y_0 = 0.189$ to zero, $\Delta A$ and $l$ both have two branches.  In this interval, the branch where $\Delta A$ is positive exist only for $l>0.5$.  In figure \ref{fig:ads4tla}, we show snapshots of $\Delta A (l)$ as $y_0$ decreases in the interval $0.217 > y_0 > 0.189$.  We observe that $\Delta A$ has a first order phase transition without changing signs; then when $y_0 <0.189$, it experiences a discontinuity from some negative value to $0$.  This phase structure is observed in the thermodynamics of charged $AdS$ black holes \cite{Chamblin:1999tk}, fundamental strings in supergravity backgrounds dual to field theories with fundamental flavor \cite{Bigazzi:2008gd} and Van der Waals gas.  In upcoming work in collaboration with C. N\'u\~nez we will elaborate on this analogy further. Here we conjecture that  $\Delta A$ behaves as a Gibbs potential or free energy depending on which cycle we fix.  In this picture, $l$ behaves like inverse temperature.

From equations (\ref{rtac}) and (\ref{rtly_s}) we find the following ultraviolet behavior for the area:
\begin{eqnarray}
A_c(l) &=& 2 \pi L^2 \left( y_\infty - \left[\frac{\sqrt{\pi}\Gamma(\frac{3}{4})}{\Gamma(\frac{1}{4})}\right]^2 \frac{2}{\tilde{\beta} l} \right) \\
&=& 2 \pi L r_\infty - \left[\frac{\sqrt{\pi}\Gamma(\frac{3}{4})}{\Gamma(\frac{1}{4})}\right]^2 \frac{4 \pi L^3}{\beta l}.
\end{eqnarray} The entropy of the smooth surface for small $l$ is then

\begin{equation}
S_c(l) = \frac{\Omega_7}{(2\pi)^6} \frac{(32 \pi^2 N)^{4/3}}{l_p} r_\infty - \left[\frac{\sqrt{\pi}\Gamma(\frac{3}{4})}{\Gamma(\frac{1}{4})}\right]^2 \frac{\Omega_7}{2 (2 \pi)^6} \frac{(32 \pi^2 N)^{5/3} l_p}{ \beta l}.
\end{equation} The entropy from the piece-wise smooth surface at small $\beta$ is
\begin{equation}
S_p = \frac{\Omega_7}{(2\pi)^6} \frac{(32 \pi^2 N)^{4/3}}{l_p} r_\infty - C \frac{\Omega_7}{ (2\pi)^5} \frac{(32 \pi^2 N)^{5/3} l_p}{ \beta }.
\end{equation}

 The ultraviolet behavior is linearly divergent, as in the previous embedding. The power of $N^{4/3}$ differs from the $N^{7/6}$ of the previous subsection and we note this as the result of breaking the conformal invariance. Namely, the presence of $r_\infty$ explicitly indicates a missing factor of $N^{1/6}$, indeed, the conformal behavior would have been observed otherwise since $N^{3/2}=N^{4/3}N^{1/6}$. As expected, the divergent piece is proportional to the area  of the codimension 3 boundary. This boundary has length $2\pi r_\infty$.  To leading order, the finite piece is proportional to $1/l$.

\subsection{General Observations}

From the above analysis, we demonstrate that the holographic geometric entropy  always indicates a transition for configurations in the Schwarzschild black hole in $AdS_4$. We found that for the minimal surface parametrized by $r({\theta})$ a transition occurs for $y_0 >0$, that is, for any nonzero temperature.  We proved this by showing that there exist a $y_*^c$ where $\Delta A$ changes sign.  It was then observed that at $y_*^c$ the value of the length is in the appropriate range:  $l<1$.  This implies that there is a $l^c$, namely $l^c=l(y_*^c)$, where $\Delta A(l)$ changes sign.

For the $r(\tau)$ embedding, we observe a more interesting structure.  The difference in area have three phases enumerated below.
\begin{enumerate}
	\item In the interval $0.217 <y_0<\infty$, $\Delta A$ is negative, and $l$ is bounded from above by $0.5$.  The result here is similar to the AdS$_3$.
	\item In the interval $0.217  > y_0 >0.189 $, $\Delta A$ and $l$ has 3 branches.  With this range, $\Delta A$ can change sign.
	\item In the interval $0.189> y_0 > 0$, $\Delta A$ and $l$ has 2 branches and $\Delta A$ is positive only for $l>0.5$.  This suggests that $\Delta A$ must jump from some negative value to zero at $l=0.5$ since the embedding exist only for $l<0.5$.
\end{enumerate}
Although we have found the critical values of the dimensionless radius of the black hole horizon $y_0^{crit1}$ and $y_0^{crit2}$ numerically, it is useful to determine them more systematically. The first critical value, $y_0^{crit1}$, occurs when $l$ and $\Delta A$ are no longer monotonic. This implies these functions have local minima and maxima.  Thus the critical value is the maximum of $y_0$ from which one starts to see these local extrema. At such value, the equation
\begin{equation}
\frac{d l}{dy_*} =0
\end{equation} must have a real solution.  At  $y_0^{crit1}$, $l(y_*)$ must also have an inflection point.  We must also have
\begin{equation}
\frac{ d^2 l}{d y_*^2} =0.
\end{equation}
So at the first critical value $y_{0}^{crit1}$, is given by the solution of the following equations
\begin{equation}
\frac{d l}{dy_*} =0, \;\;\; \mbox{and} \;\;\; \frac{ d^2 l}{d y_*^2} =0.
\end{equation}
The second critical value, occurs when $l(y_*)$ transitions from having 3 branches to 2.  This occurs when one of the zeros of $\frac{d l}{dy_*}$ is $y_{0}^{crit2}$.  Thus at the second critical point, we have:
\begin{equation}
\left. \frac{d l}{dy_*}\right|_{(y_*=y_0^{crit2})} =0.
\end{equation}

\section{Geometric entropy from $AdS_5$} \label{ads5}

Consider the Schwarzschild black hole in global $AdS_5$ with the following metric
\begin{equation}
ds^2 = f(r) d\tau^2  + \frac{1}{f(r)} dr^2 + r^2 \left( d\theta^2 + \sin^{2}(\theta)(d\psi^2 + \sin^{2}(\psi)d\phi^2)\right)
\end{equation} with
\begin{equation}
f(r) = 1 - \frac{M}{r^2} + \frac{r^2}{L^2}.
\end{equation}
Introducing $y=r/L$ and $m=M/L^2$, we have
\begin{equation}
f(y) = 1 -\frac{m}{y^2} + y^2 .
\end{equation}
We consider the entropy when $\phi$ is fixed. The periodicity in $\tau$ is set to $\beta$ to avoid conical singularities in that direction, its dimensionless periodicity is
\begin{equation}
\tilde{\beta} = \frac{\beta }{L} = \frac{4 \pi y_0}{2+4y_0^2}.
\end{equation}

\subsection{r($\psi$) Minimal surface}

First we evaluate the smooth minimal surface. The induced metric is
\begin{equation}
ds^2 = f(r) d\tau^2 + (r^2 b^2 + r^2 \sin^2(\theta)) d\psi^2 + r^2 d\theta^2, \;\;\; \mbox{with} \;\;\;  b = \frac{r'(\psi)}{r \sqrt{f(r)}}.
\end{equation}
The area is then
\begin{equation}
A_c = \beta \int r^2 \sqrt{f(r)}\sqrt{b^2 + \sin^2(\theta)} d\theta d\psi = \beta \int r^2 \sqrt{f(r)} p(b) d\psi, \quad 
p(b) = \int_0^\pi \sqrt{b^2 + \sin^2(\theta)} d\theta,
\end{equation} 
where $p(b)$ is an elliptic function.
The equation of motion is
\begin{equation}
r_*^2 \sqrt{f(r_*)} q(0) =r^2 \sqrt{f(r)} q(b), \;\;\; \mbox{with} \;\;\; q(b) = \int_0^\pi \frac{\sin^2(\theta)}{\sqrt{b^2 + \sin^2(\theta)}} d\theta.
\end{equation}
The value of $r_*$ corresponds to $r' =0$ or $b=0$.  After switching integration from $\psi$ to $r$, we obtain
\begin{equation}
A_c = 2\beta \int_{r_*}^{r_\infty} \frac{r}{b} p(b) dr. 
\end{equation}
The piece-wise smooth surface evaluates to:
\begin{equation}
A_p = 2\beta \int_{r_0}^{r_\infty} r dr \pi = \pi \beta (r_\infty^2 - r_0^2)
\end{equation}
The difference in areas is
\begin{equation}
\Delta A = 2 \beta \int_{r_*}^{r_\infty} \frac{r}{b}(p(b) - \pi b) dr + \beta \pi (r_0^2 - r_*^2).
\end{equation}
Now we want to set up the problem so that we can integrate over $b$.  The equation of motion is not so readily solvable to give $r(b,r_*)$ due to the presence of $f(r)$.  In the $y$ coordinate, the equation of motion can be written as
\begin{equation}
y^6 + y^4 -m y^2 -\rho_*^4 g^2(b) =0  \label{eom}
\end{equation}
where
\begin{equation}
\rho_*^2 = y_*^2 \sqrt{f(y_*)},\;\;\; m= y_0^4 + y_0^2 \;\;\; \mbox{and} \;\;\; g(b) = \frac{q(0)}{q(b)}.
\end{equation}
As integrals over $b$, $l$ and $\Delta A$ are
\begin{eqnarray}
\Delta A &=& 2 L^2 \beta \int_0^\infty \frac{d y(y_*,b)}{db} \frac{y(y_*,b)}{b} (p(b) - \pi b) db + L^2 \beta  \pi (y_0^2 - y_*^2), \\
l &=& \frac{2}{\pi} \int_0^\infty \frac{d y(y_*,b)}{db} \frac{db}{b \sqrt{y^4 + y^2 -m}} = \frac{2}{\pi} \int_0^\infty y \frac{dy}{db}\frac{db}{b g(b)}.
\end{eqnarray} One solves for $y(y_*,b)$ from (\ref{eom}) and then evaluates these integrals.  Furthermore, by using (\ref{eom}), one can write

\begin{equation}
\frac{dy}{db} = \frac{\rho_*^4 g(b)}{y} \frac{dg(b)}{db} \frac{1}{3y^4 + 2y^2 -y_0^4 -y_0^2}
\end{equation} from which one obtains,

\begin{eqnarray}
\Delta A &=& 2 L^2 \beta \int_0^\infty \frac{\rho_*^4 g(b)}{b} \frac{dg(b)}{db} \frac{1}{3y^4 + 2y^2 -y_0^4 -y_0^2} (p(b) - \pi b) db + L^2 \beta  \pi (y_0^2 - y_*^2) \nonumber \\
l &=& \frac{2\rho_*^2}{\pi} \int_0^\infty \frac{1}{b} \frac{dg(b)}{db} \frac{db}{3y^4 + 2y^2 -y_0^4 -y_0^2}.
\end{eqnarray}
As in the case case of $AdS_4$, we explore two regimes determined by $y_*$, that is by the turning point of the surface. For $y_*\to y_0$,
it can be shown that
\begin{equation}
 \Delta A(y_0 + \epsilon) > 0.
\end{equation}
In the other limit, that is, for large $y_*$, we have $\rho_*^4 \approx y_*^6$ and $y^4 \approx y_*^4 g^{4/3}(b)$.  This allows us to write $\Delta A$, in the large $y_*$ limit, as
\begin{eqnarray}
\Delta A &=& 2 L^2 \beta \int_0^\infty \frac{y_*^6 g(b)}{b} \frac{dg(b)}{db} \frac{1}{3y^4} (p(b) - \pi b) db - L^2 \pi \beta  y_*^2 \\
&=& 2 L^2 \beta y_*^2 \left[\int_0^\infty \frac{ g(b)}{b} \frac{dg(b)}{db} \frac{1}{3 g^{4/3}(b)} (p(b) - \pi b) db - \frac{\pi}{2} \right] < 0
\end{eqnarray}
Since
\begin{equation}
 \int_0^\infty \frac{ g(b)}{b} \frac{dg(b)}{db} \frac{1}{3 g^{4/3}(b)} (p(b) - \pi b) db = 0.9172\cdots .
\end{equation}
This shows that $\Delta A$ changes sign for some ranges of $y_0$. Similar to $r(\theta)$ in $AdS_4$, this is not enough to conclude that the difference in area generically changes sign.  We need to show that $l$ is appropriately bounded.

Let us present the ultraviolet behavior of the entropy. In the large $y_*$ limit, we can write,
\begin{eqnarray}
A_c &=& 2 L^3 \tilde{\beta} \int_0^\infty \frac{y_*^6 g(b)}{b} \frac{dg(b)}{db} \frac{1}{3y^4} p(b) db, \;\;\; y^4 \approx y_*^4 g^{4/3}(b) \\
&=& \tilde{\beta} L^3 \pi y_\infty^2 - L^3 \tilde{\beta} y_*^2 1.307...
\end{eqnarray}
Using the expression for $l$ in this limit
\begin{eqnarray}
l &=& \frac{2}{\pi y_*} \int_{0}^{\infty} \frac{dg}{db} \frac{1}{3 b g^{4/3}} db = \frac{0.416}{y_*},
\end{eqnarray}
we  obtain
\begin{eqnarray}
A_c(l) &=& \beta \pi r_\infty^2 - L^2 \beta \frac{21.8}{\pi^2 l^2}.
\end{eqnarray}
We can write the entanglement entropy, $S = \frac{A}{4 G_N^{(5)}}$, in terms of field theory parameters from the relations
\begin{equation}
G_N^{(10)} = 8 \pi^6 g_s^2 \alpha'^4 = 8 \pi^6 \frac{\alpha'^4 \lambda^2}{N_c^2}, \;\;\; L^4 = 4 \pi \alpha'^2 \lambda, \;\;\; G_N^{(5)} = \frac{G_N^{(10)}}{L^5 \Omega_5}.
\end{equation}

The entropy from the smooth surface for small $l$ is

\begin{equation}
S_c(l) =  \frac{(4 \pi)^{5/2} \Omega_5}{32 \pi^5} \frac{1}{\l^{3/4}}\,\,\, N_c^2 \,\, \frac{\beta r_\infty^2}{l_s^3}  - \frac{(4 \pi)^{7/4} \Omega_5 }{32 \pi^6} \frac{N_c^2}{\lambda^{1/4}}  \frac{21.8 \beta }{\pi^2 l_s }\frac{1}{l^2}.
\end{equation}
where $l_s$ is the string length scale $l_s=\sqrt{\alpha'}$. We note a few interesting general properties that we reproduce. First, there is the quadratic divergence that is typical to four-dimensional field theories. For example, Srednicki had already discussed it in \cite{Srednicki:1993im}.  It was also more recently discussed and obtained for various field theories, including their holographic formulation by Ryu and Takayanagi in \cite{Ryu:2006ef}. The $1/l^2$ dependence of the finite piece is also rather general in four dimensional field theories, as widely discussed in \cite{Ryu:2006ef}. Both these statements seems to follow from dimensional analysis but it interesting that they are confirmed by the holographic calculations. As expected, we also obtain the $N_c^2$ factor in the expression for the entropy. However, unexpectedly we have powers of the 't Hooft coupling $\l$. We explained these powers as effects of breaking conformal invariance by introducing some scales in the problem, in this case $\beta$. Recall that in the standard holographic calculation of the entanglement entropy there are no powers of $\lambda$ \cite{Ryu:2006bv,Ryu:2006ef}. We find this dependence on $\lambda$ very peculiar and it would be interesting to pursue its interpretation.
We can also write the entropy coming from the piece-wise smooth surface.  This entropy is independent of $l$ and is given as
\begin{equation}
S_p =  \frac{(4 \pi)^{5/2} \Omega_5}{32 \pi^5} \frac{1}{\l^{3/4}}\,\,\, N_c^2 \,\, \frac{\beta r_\infty^2}{l_s^3} - \frac{(4\pi)^{9/4} \Omega_5}{128 \pi^4 } \lambda^{1/4}\frac{l_s}{\beta} \,\,\,N_c^2 \left(1 + \left(1-\frac{2 \beta^2}{\pi^2 \sqrt{4\pi \alpha'^2 \lambda}}\right)^{1/2} \right)^2.
\end{equation}  Here we have to replace $r_0$ for $\beta$ through the relationship,
\begin{equation}
r_0 = \frac{\pi L^2}{2 \beta} \left(1 + \left(1 - \frac{2\beta^2}{\pi^2 L^2}\right)^{1/2} \right),
\end{equation}
for a thermodynamically stable black hole.  Notice that there is a maximum $\beta$ for which the black hole exist.

\begin{figure}[ht]
\centering
	\subfigure[]{\label{fig:ads5lpsi}\includegraphics[scale=.4]{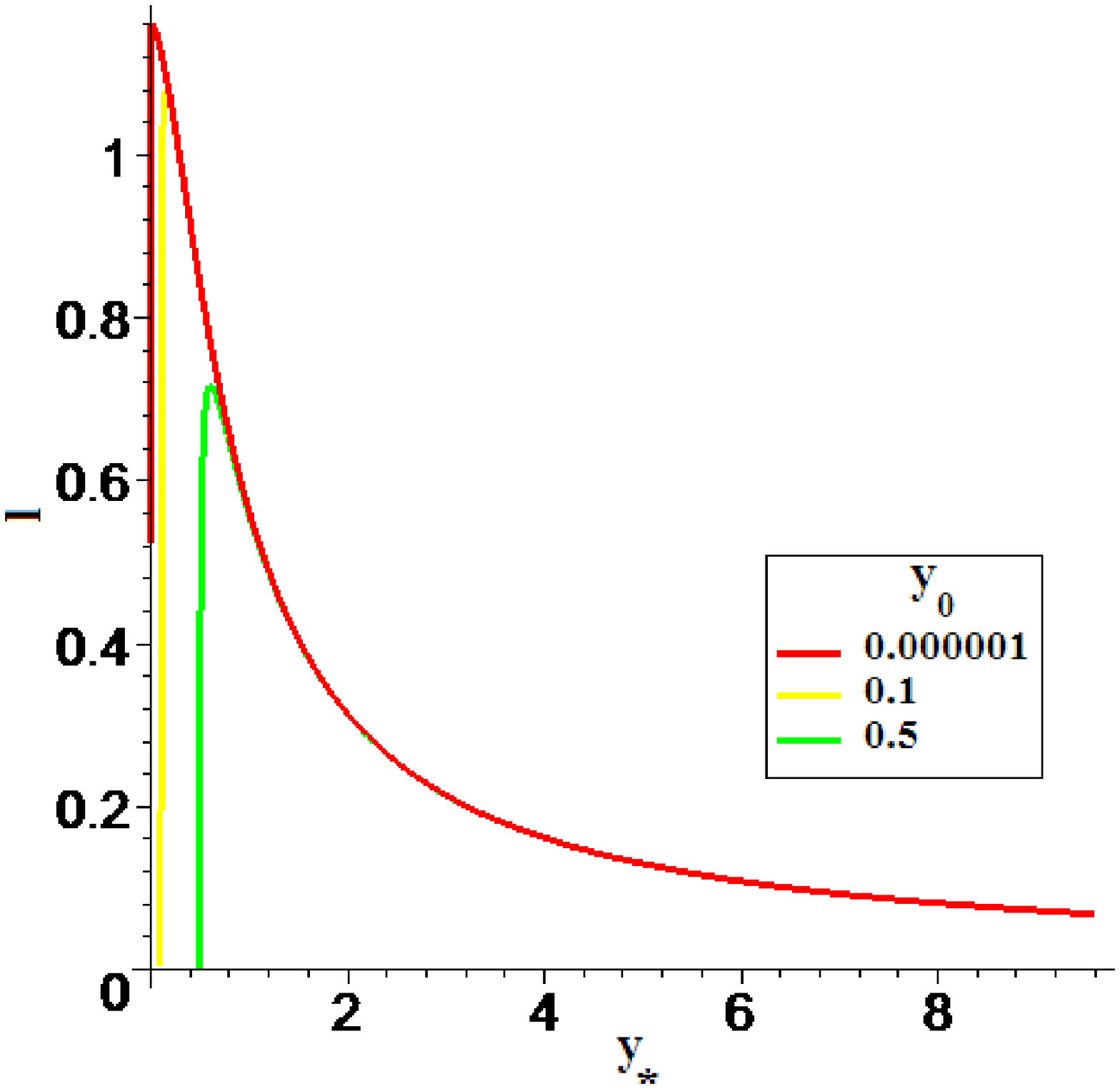}}
	\subfigure[]{\label{fig:ads5apsi}\includegraphics[scale=.4]{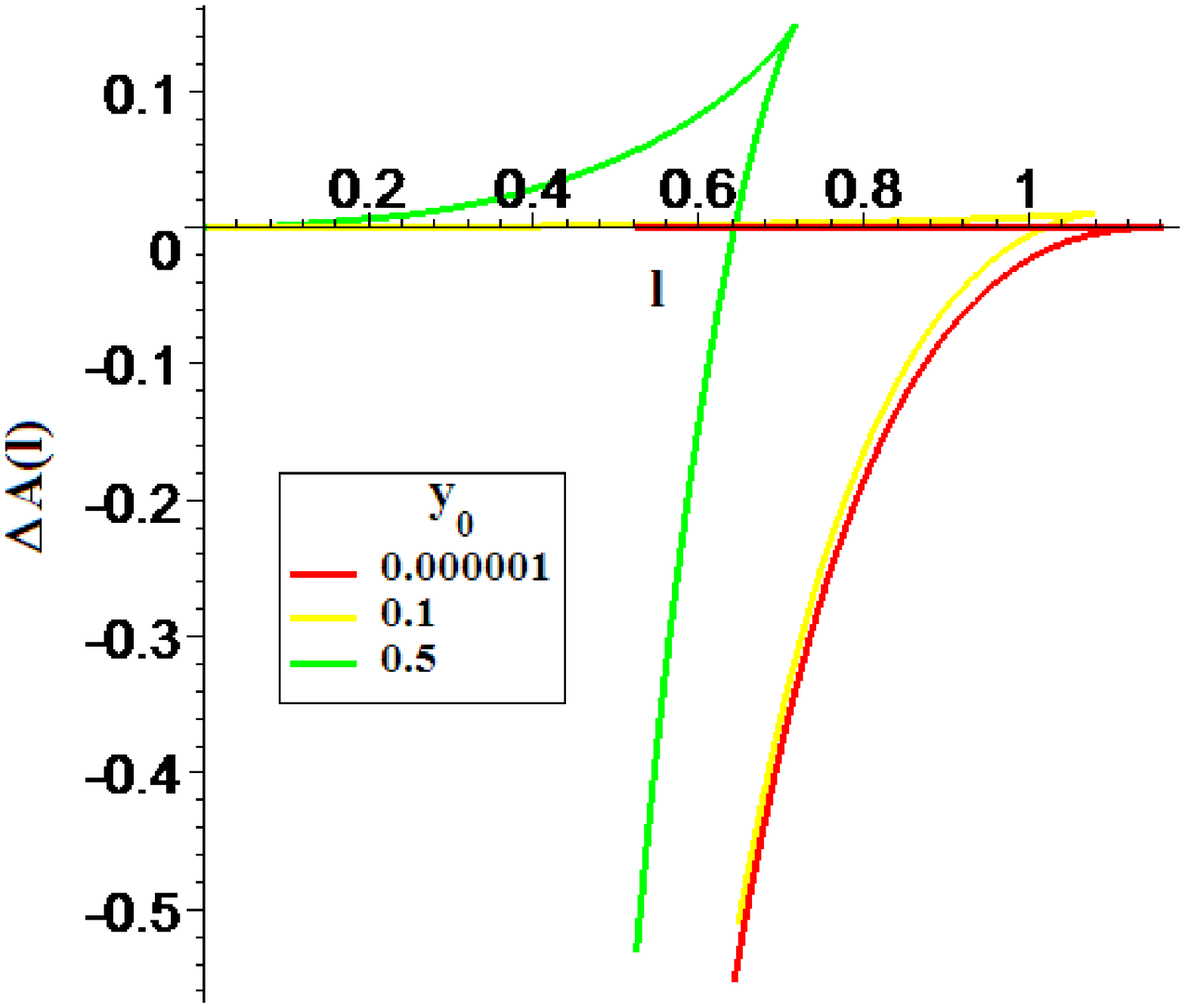}}
 	\caption{\ref{fig:ads5lpsi} shows the function $l(y_*)$ at $y_0 = 10^{-6}, 0.1, 0.5$.   \ref{fig:ads5apsi} shows the behavior of $\Delta A$ as a function of $l$.  We generically observe a transition.}\label{ads5psi}
\end{figure}

We plot some results that verify numerically that we always get a transition. We note, as we did in the introduction and outlook section, that the result is qualitatively the same as for the $r(\theta)$ embedding in $AdS_4$.

\subsubsection{$r(\tau)$ Minimal surface}
\begin{figure}[h]
\centering
		\subfigure[$y_0 = .3$]{\label{fig:ads5tla-a}\includegraphics[scale=.5]{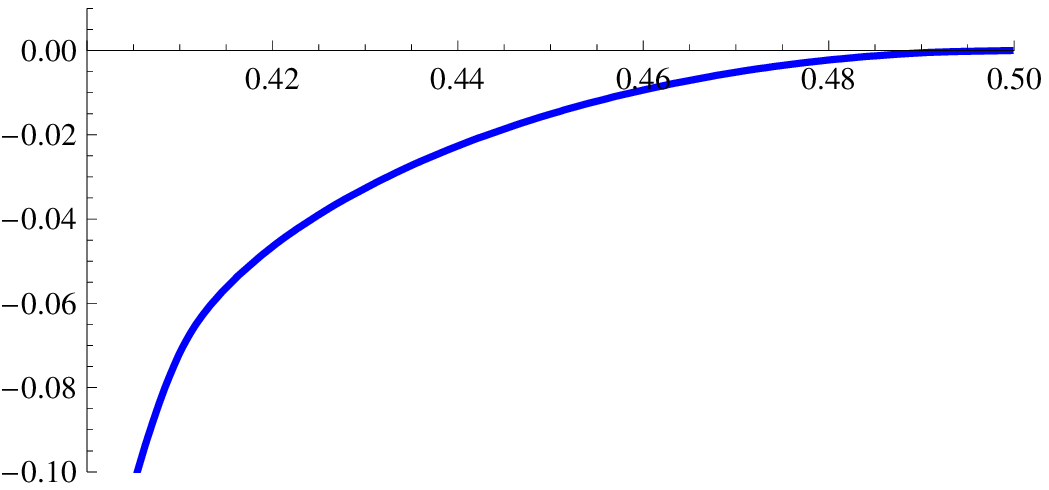}}
		\subfigure[$y_0 = .286$]{\label{fig:ads5tla-b}\includegraphics[scale=.5]{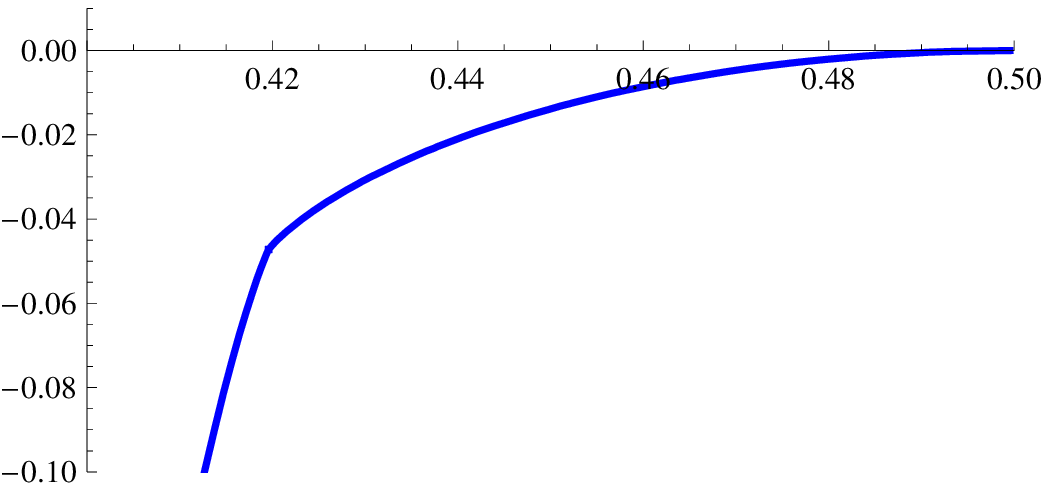}}	 	  	
		\subfigure[$y_0 = .266$]{\label{fig:ads5tla-c}\includegraphics[scale=.5]{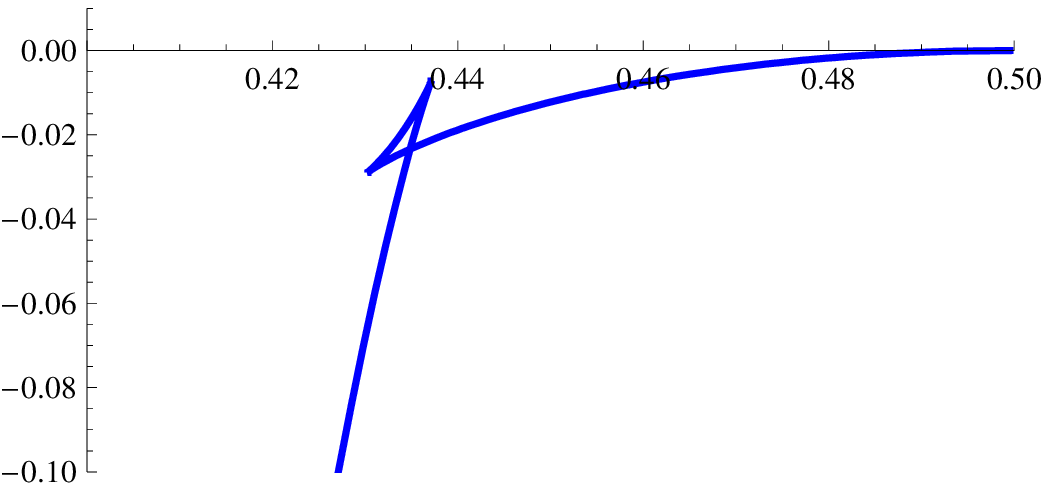}}
		\subfigure[$y_0 = .25$]{\label{fig:ads5tla-d}\includegraphics[scale=.5]{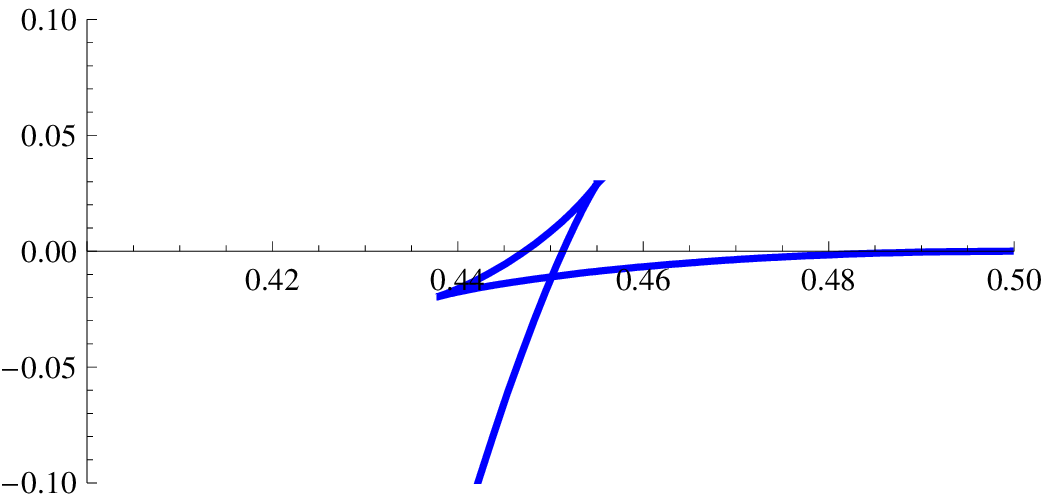}}	
		\subfigure[$y_0 = .23$]{\label{fig:ads5tla-e}\includegraphics[scale=.5]{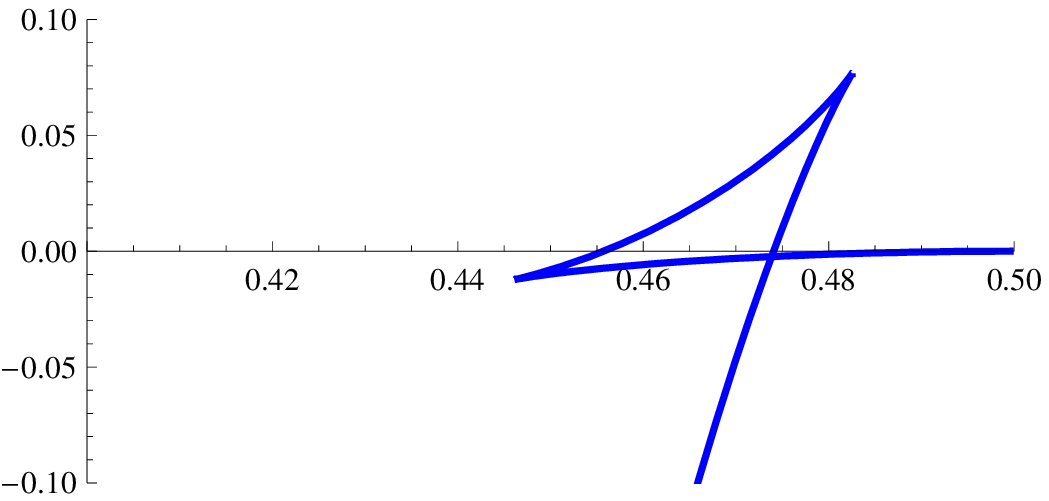}}	
		\subfigure[$y_0 = .22$]{\label{fig:ads5tla-f}\includegraphics[scale=.5]{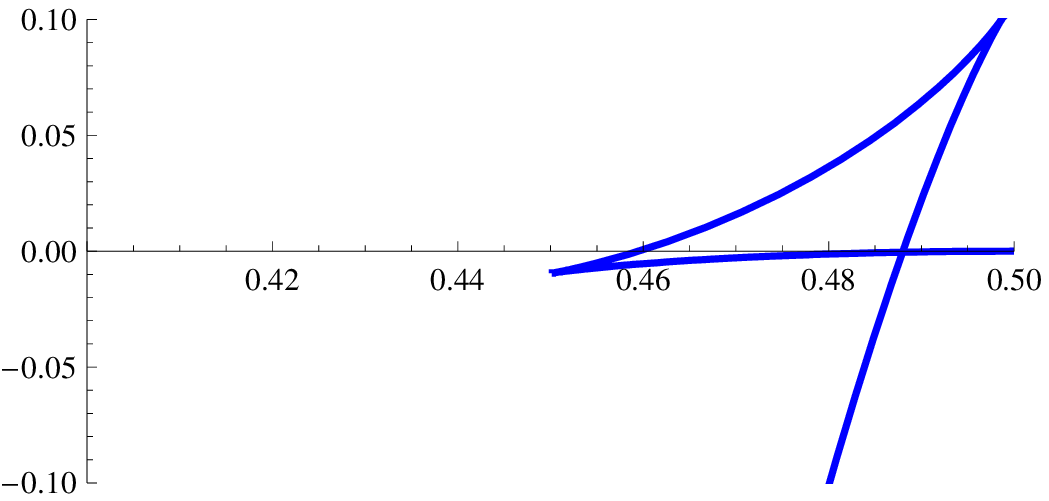}}	
		\subfigure[$y_0 = .212$]{\label{fig:ads5tla-g}\includegraphics[scale=.5]{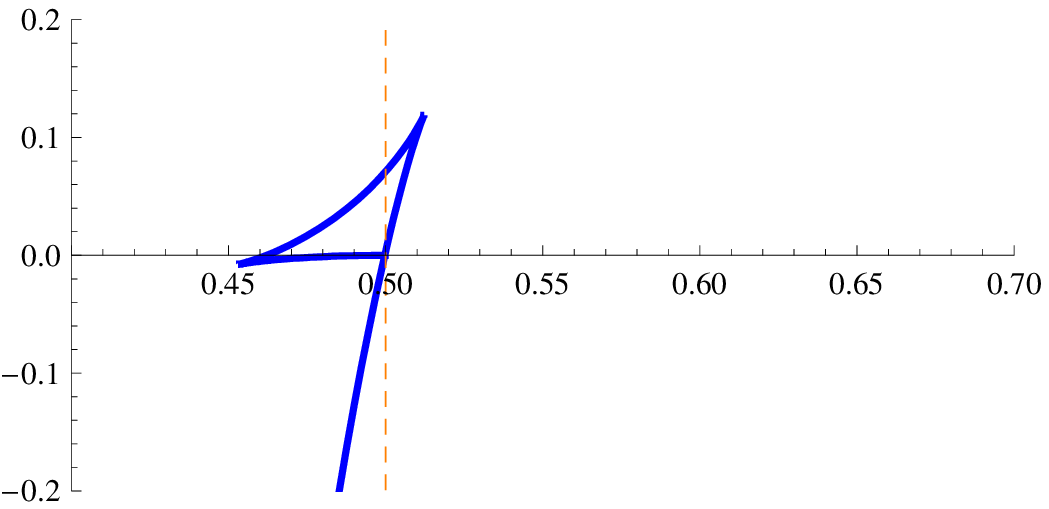}}	
		\subfigure[$y_0 = .19$]{\label{fig:ads5tla-h}\includegraphics[scale=.5]{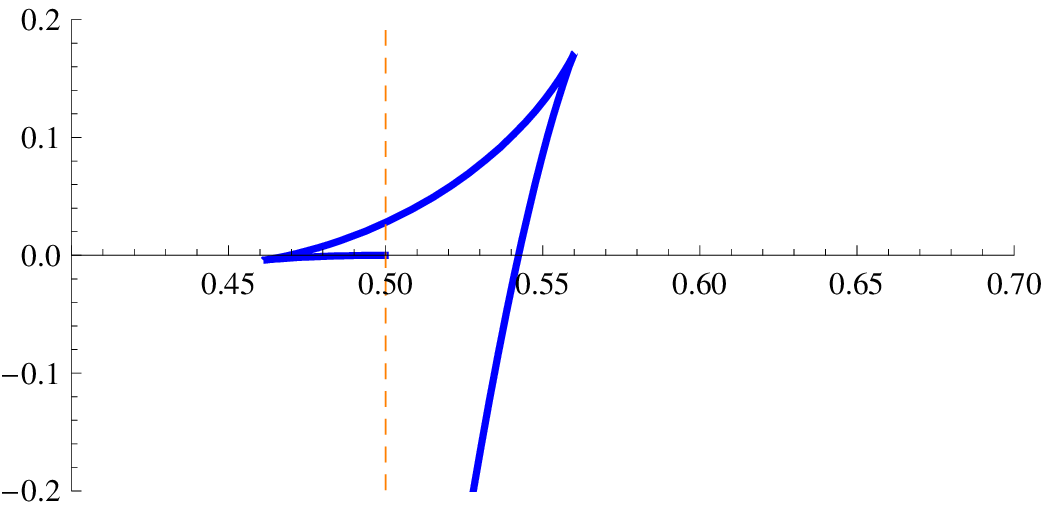}}	
		\subfigure[$y_0 = .01$]{\label{fig:ads5tla-i}\includegraphics[scale=.5]{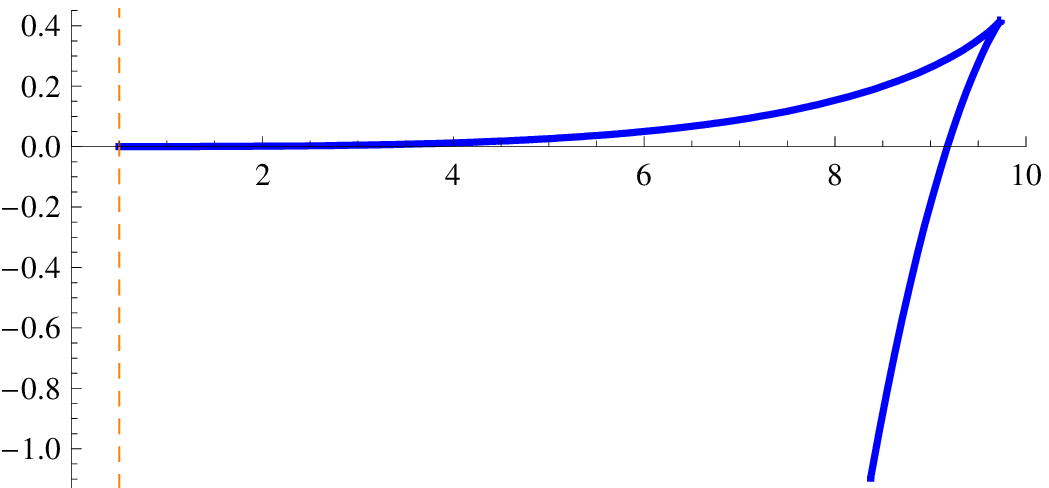}}	
												
	\caption{We plot the difference in entropy for $r(\tau)$ in $AdS_5$ vs. $l$ at difference values of $y_0$.  These snapshots track the the difference in entropy as it experiences a first order phase transition.}
	\label{ads5tla}
\end{figure}

The induced surface in the Euclidean frame is given as

\begin{equation}
ds^2 = f(r)(1 + \frac{r'^2(\tau)}{f^2(r)}) d\tau^2 + r^2 d\Omega_2^2 .
\end{equation}
The subsystem is given by the segment $0\leq \tau \leq \beta l$, the area is given as
\begin{equation}
A = 2\pi \int_{0}^{\beta l} r^2 \sqrt{f(r)}\sqrt{1 + \frac{r'^2(\tau)}{f^2(r)}} d\tau,
\end{equation}
where prime is derivative with respect to $\tau$. The equation of motion is then
\begin{equation}
\frac{r'^2(\tau)}{f^2(r)} = \frac{r^4 f(r)}{r_*^4 f(r_*)} -1.
\end{equation}
The area of the surface can then be written as
\begin{eqnarray}
A &=& 2 \pi \int_0^{\beta l} \frac{r^4 f(r)}{r_*^2 \sqrt{f(r_*)}} d\tau = 8 \pi \int_{r_*}^{r_\infty}  \frac{ r^4}{ \sqrt{ r^4 f(r) -r_*^4f(r_*)}} dr .
\end{eqnarray}
The piece-wise smooth surface is given by $\tau=$constant, with area
\begin{eqnarray}
A_p = 4 \pi \int_{r_0}^{r_{\infty}} \frac{r^2}{\sqrt{f(r)}} dr &=& 2 \pi L r_\infty^2 - \pi L^3 \ln \left(  \frac{4 r_\infty^2}{L^2 + 2 r_0^2} \right). \label{Apads5}
\end{eqnarray}
The difference in area is then
\begin{eqnarray}
\Delta A &=& 4 \pi L^3 \left[ \int_{y_*}^{y_\infty}  \left(\frac{ y^4}{ \sqrt{ y^4 f(y) -y_*^4f(y_*)}} - \frac{y^2}{\sqrt{f(y)}} \right) dy - \int_{y_0}^{y_{*}} \frac{y^2}{\sqrt{f(y)}} dy \right] .
\end{eqnarray}
We can also write the relationship $l(y_*)$,
\begin{equation}
l = \frac{2L}{\beta} y_*^2 \sqrt{f(y_*)} \int_{y_*}^{y_\infty} \frac{dy}{f(y)\sqrt{ y^4 f(y) -y_*^4f(y_*)}}.
\end{equation}
Let us now consider the limits of $\Delta A$. Similar to the $AdS_4$ case, we immediately obtain
\begin{equation}
\Delta A (y_* =y_0) = 0.
\end{equation}
In the case when $y_*$ is large, we can rewrite the integral as
\begin{eqnarray}
\Delta A &=& 4 \pi L^3 \left[ y_*^2 \int_{1}^{x_\infty}  \left(\frac{ x^4}{ \sqrt{ x^6 -1}} - x \right) dx - \int_{y_0}^{y_{*}} \frac{y^2}{\sqrt{f(y)}} dy \right] \\
&=& 4 \pi L^3 \left[0.28440 y_*^2 - \int_{y_0}^{y_{*}} \frac{y^2}{\sqrt{f(y)}} dy \right] < 0.
\end{eqnarray}
In the first integral, we used the fact that for large $y$, $f(y) \approx y^2$ and then introduced the coordinate $x=y/y_*$.  The inequality in the last line follows from the fact that for large $y_*$ the second integral diverges as $\frac{1}{2}y_*^2$.  The change in the difference of areas from zero to negative does not, by itself, indicates a transition. A similar situation was observed in the case of  $AdS_4$. Although more work is still required at this point, we verify explicitly that the result for $AdS_5$ is qualitatively the same as in $AdS_4$, the difference is in the actual numerical values of the critical temperatures, that is, of the critical values of $y_0$. In figure \ref{ads5tla}, we plot snapshots of $\Delta A (l)$ in the range of $y_0$ where we observe first order phase transitions.

The ultraviolet behavior of the entropy can be obtained as before.
In the large $y_*$ limit, we need the large $y$ limit of $f(y)$, $f(y) \approx y^2$.  We thus obtain
\begin{eqnarray}
A_c &=& 2\pi L^3 y_\infty^2 - 2 \pi L^3 \ln(y_\infty) - 4\pi L^3 y_*^2 0.216, \qquad
l = \frac{2}{\tilde{\beta} y_*} 0.431,
\end{eqnarray}
which translates into the following result
\begin{eqnarray}
A_c(l) &=& 2\pi L r_\infty^2 - 2 \pi L^3 \ln\left(\frac{r_\infty}{L}\right) - 4 \pi L^5 \frac{0.160}{\beta^2 l^2}.
\end{eqnarray}
We can write the entanglement entropy, $S = \frac{A}{4 G_N^{(5)}}$, in the small $l$ for the smooth surface as
\begin{eqnarray}
S_c(l) &=& \frac{(4\pi)^{3/2} \Omega_5}{16 \pi^5} \frac{N_c^2}{\lambda^{1/2}} \frac{r_\infty^2}{l_s^2} - \frac{  \Omega_5}{ \pi^3} N_c^2 \ln\left(\frac{r_\infty}{(4 \pi \lambda)^{1/4}l_s}\right) \nonumber \\
&-&  \frac{(4\pi)^{5/2} \Omega_5}{ 8 \pi^5}\frac{l_s^2}{\beta^2}   \frac{0.16 \lambda^{1/2}N_c^2}{ l^2}.
\end{eqnarray}
The entropy of the piece-wise smooth surface can be obtained from (\ref{Apads5}), it is given by
\begin{eqnarray}
S_p &=& \frac{(4\pi)^{3/2} \Omega_5}{16 \pi^5} \frac{N_c^2}{\lambda^{1/2}} \frac{r_\infty^2}{l_s^2} - \frac{  \Omega_5}{ \pi^3} N_c^2 \ln\left(\frac{r_\infty}{( \pi \lambda)^{1/4}l_s}\right) \\
&+&  \frac{  \Omega_5}{ 2 \pi^3} N_c^2 \left[ \ln\left(\frac{\pi^2 (4 \pi  \lambda)^{1/2}l_s^2}{2 \beta^2}\right) + \ln \left(1+\left(1-\frac{2 \beta^2}{\pi^2 (4 \pi \lambda)^{1/2}l_s^2}\right)^{1/2}\right) \right]. \nonumber
\end{eqnarray}

In addition to reproducing the area law, we observe that the divergence here have a logarithmic term.

\section{Comments on the various cycles}\label{sec:hp}
In the previous sections we have considered various configurations where we have fixed a periodic coordinate $\phi$. In previous work \cite{Faraggi:2007fu} we considered fixing the temporal cycle which we denoted by $\tau$ in this paper. A natural question then arises as to what is the relationship between these two cycles and what are the rules for choosing one or the other. 

It is of course, a well known fact that the presence of two potential time cycles usually induces transitions in field theory. Generically there is a low temperature phase and a high temperature phase. Perhaps the cleanest example is Witten's reinterpretation of the Hawking-Page phase transitions \cite{Hawking:1982dh,Witten:1998zw}. There the competition between two saddle points of the action, namely, the Schwarzschild black hole in $AdS$ and the thermal $AdS$ was interpreted as the supergravity dual of the confinement/deconfinement phase transition for ${\cal N}=4$ supersymmetric Yang-Mills on the sphere. A slightly more complicated version was discussed in \cite{Aharony:2006da}, where the transition between the low temperature and high temperature phase was shown explicitly by evaluating the actions of various compactifications of D4 branes. In the context of supergravity backgrounds dual to confining field theories the analog of the Hawking-Page phase transition was observed numerically in \cite{Mahato:2007zm}, similar results were reported in \cite{Aharony:2007vg}. In every case the dominant thermal phase is determined by comparing the supergravity actions.

The problem we have at hand, that is, the competition between the backgrounds where temporal boundary conditions are imposed along the $\tau$ cycle and along the $\phi$ cycle seems to have the same flavor. However, a naive evaluation of the difference of actions, which in our case, as in the case of \cite{Witten:1998zw}, boils down to an evaluation of the difference in volumes yields a very puzzling result. Namely, if we evaluate the difference in actions from the manifold with temporal boundary conditions along $\tau$ and along $\phi$ we obtain a result of the form:
\be
\Delta V=V_{\tau}-V_{\phi} \sim \frac{\beta_0(r_{\infty}^2-r_+^2)}{2}\left(2\pi -\frac{\beta_0}{L} \right),
\ee
where $\beta_0$ is the black hole  inverse temperature. This result is puzzling because, since $\Delta S\sim \Delta V$,  it seems to suggest that for high temperature the background (or the phase) that dominates is the one with temporal conditions along $\phi$. Our intuition tells otherwise, we believe that the high temperature phase that dominates is the Schwarzschild black hole.
We do not have a full resolution of this puzzle and hope to return to it in a separate publication. However, there are a few things that might go wrong with the above naive evaluation. For instance, the conical singularity along the $\phi$ cycle prevents us from evaluating the action simply as a volume since the curvature will develop singularities. 

We finish this section mentioning another idea that evolved largely from our previous work \cite{Faraggi:2007fu} and correspondence with T. Takayanagi. Namely, that the calculations presented in this paper, which fully involved the $\tau$-cycle, are to be understood as the Polyakov loop type of generalization of the standard entanglement entropy. This is a thought that we also plan to develop separately but that seems worth mentioning given the substantial difference between the phase structure observed in \cite{Faraggi:2007fu} with generically no phase transitions and the rich structure encountered here.

\section{Conclusions}\label{conclusions}
In this paper we have investigated the holographic proposal for the geometric entropy by computing the entanglement entropy for Schwarzschild black holes in global $AdS_p$ spaces.
In the Euclidean version of the gravity geometry we have various choices of the subsystem $A$. 
Our calculations show,
as already suggested in \cite{Fujita:2008zv}, that the geometric entropy we compute here is the analog of the Polyakov loop in field theory, whereas the calculations of the entanglement entropy in \cite{Faraggi:2007fu} are more like the analog of the calculations of the Wilson loop.
 
For $AdS_3$ we found a perfect agreement with the form of the geometric entropy computed for a two-dimensional CFT with periodic boundary conditions.  
We have also found 
(and believe that it is one remarkable result of our investigation) 
an universality of the phase diagrams for both $AdS_4$ and $AdS_5$. 

When interpreted from the field theory point of view our results fits many expectations about the structure of divergences and the various dependence of the entanglement entropy. In particular, we found an agreement with various expectations in $2+1$ dimensions \cite{Kitaev:2005dm} and \cite{Pakman:2008ui}. Similarly, our results fit nicely with expectation for $3+1$ dimensional field theories as presented in \cite{Srednicki:1993im,Ryu:2006ef}. However, due to the fact that we introduce various scales in the problem, our results have different $N$ scaling than the expected from a conformal theory. We discussed this in the context of the dual field theories. We believe it would be interesting to pursue the implications of these scalings as potential phases that have not been considered before. In a sense this situation is similar to generalizations of the UV/IR relation in the context of the AdS/CFT. It would be interesting to find,  as in \cite{Horowitz:1999gf,Danielsson:1998wt},  precise descriptions on both sides of the correspondence of probes which yield different energy-in-the-boundary/distance-in-the-bulk relations.


One interesting problem that we also plan to pursue is the following.
From the field theory point of view our calculation corresponds to the geometric entropy for a field theory  in a thermal state $|\Psi\rangle$.  In much of the literature the state $|\Psi\rangle$ has been taken to be the ground state. What we found out in the present paper and in our previous explorations \cite{Faraggi:2007fu}, is that not only the state $|\Psi\rangle$ is crucial but also that the region $A$ plays a central role. It would be interesting to compute the geometric entropy at finite temperature directly in field theory.

\section*{Acknowledgments}
We are grateful  to R. Akhoury for various explanations concerning the entanglement entropy in QCD, to A. Hashimoto for some insightful comments and to  T. Takayanagi for correspondence. We are particularly grateful to C. N\'u\~nez for collaboration in related areas and various comments. I.B. is partially supported by a Rackham Science Award.  This work is  partially supported by Department of Energy under
grant DE-FG02-95ER40899 to the University of Michigan.

\appendix

\section{Piece-wise smooth minimal surface \label{horizon}}

We want to determine the minimal piece-wise smooth surface that is connected to the horizon.  

Given the metric 

\begin{equation}
ds^2 = \tau(r) dt^2 + h(r) dr^2 + s(r,y^i)dx^2 + \sigma_{ij}(r,y^i) dy^idy^j
\end{equation} with $r_0 \leq r <\infty$, $h(r \to r_0) \to \infty$ and $\tau(r \to r_0) \to 0$

We can parametrize the surface as:
\begin{equation}
\phi_1 = t-t_0, \;\;\;  \phi_2 = \begin{cases} x - x_a(r)  \;\;\; \infty \leq r \leq r_0, \;\; -l/2 \leq x \leq a \;\; \mbox{s.t.} \;\; x_a(r_0) = a \\ r-r_0 \;\;\; a \leq x \leq b \\ x - x_b(r)  \;\;\; r_0 \leq r \leq \infty, \;\; b \leq x \leq l/2 \;\; \mbox{s.t.} \;\; x_b(r_0) = b. \end{cases} \end{equation} $x_a$ and $x_b$ are determined by minimizing their area under the given boundary conditions.  The minimal surface connecting $x_a$ and $x_b$ at $r_0$ is given by the segment on the horizon.  We start by proving this.  Parametrize the surface as
\begin{equation}
t = \mbox{constant}, \;\;\; r=r(x) \;\;\; \mbox{with b.c.} \;\;\; r(a)=r(b) = r_0.
\end{equation} 

The induced metric is 
\begin{equation}
ds^2 = [s(r,y^i) + h(r) r'^2(x)]dx^2 + \sigma_{ij} dy^idy^j
\end{equation} with area given as
\begin{equation}
A = \int_a^b dx \int dY \sqrt{\sigma} \sqrt{s + h r'^2}
\end{equation} where $dY \sqrt{\sigma}$ is the volume element of the internal manifold.  The problem of finding the minimal surface reduce to finding the equation of motion for a point particle $r(x)$ with Lagrangian 
\begin{equation}
L = \int dY \sqrt{\sigma} \sqrt{s + h r'^2}.
\end{equation} Since this Lagrangian does not explicitly depend on $x$, the quantity 
\begin{equation}
E = \frac{\partial L}{\partial r'} r' - L 
\end{equation} is constant.  Thus the e.o.m. is 

\begin{equation}
E = - \int \frac{s\sqrt{\sigma} dY}{\sqrt{s + h r'^2}}.
\end{equation}

Since $h(r \to r_0) \to \infty$, then a nonzero value of $E$ requires $h(r(x\to a))r'^2(x \to a)=h(r(x \to b))r'^2(x \to b)=c$ where $c$ is a constant, thus $r'(a)=r'(b)=0$.  So E is simply given by
\begin{equation}
E = - \int \frac{s(r_0,y^i)\sqrt{\sigma(r_0,y^i)} dY}{\sqrt{s(r_0,y^i) + c}}
\end{equation}

Since $r(a)=r(b)=r_0$, there exist a $x_0$ in the interval $(a,b)$ where the function $r(x)$ hits a maximum value, $r_m$.  if $r(x)$ is smooth, then $r'(x_0)=0$ thus 
\begin{equation}
E =- \int \sqrt{s(r_m,y^i)\sigma(r_m,y^i)}dY. 
\end{equation} We then have 
\begin{equation}
\int \frac{s(r_0,y^i)\sqrt{\sigma(r_0,y^i)} dY}{\sqrt{s(r_0,y^i) + c}} = \int \sqrt{s(r_m,y^i)\sigma(r_m,y^i)} dY.
\end{equation} We observe that

\begin{equation}
\int \frac{s(r_0,y^i)\sqrt{\sigma(r_0,y^i)} dY}{\sqrt{s(r_0,y^i) + c}} \begin{cases} \leq \int \sqrt{s(r_0,y^i)\sigma(r_0,y^i)} dY   \;\;\; \mbox{if} \;\;\; c \geq 0 \\ \geq \int \sqrt{s(r_0,y^i)\sigma(r_0,y^i)} dY  \;\;\; \mbox{if} \;\;\; c \leq 0 \end{cases}.
\end{equation} If the quantity 

\begin{equation} 
\sqrt{H(r)} \equiv \int \sqrt{s(r,y^i)\sigma(r,y^i)} dY
\end{equation} is increasing in $r$, then $r_m \geq r_0$ exist only when $c\leq 0$.  Since $c$ cannot be negative, then $r_m$ must be equal to $r_0$.  So we observe that the minimal surface is given by $r(x)=r_0$.  

Now we determine $x_a$ and $x_b$.  In what follows, we drop the dependence of $y$ in $s(r,y^i)$.  As noted above, the $y$ dependence is irrelevant. The induced metric under these constraints is 
\begin{eqnarray} 
ds^2 &=& h^2(r) dr^2 + s^2(r)\left(\frac{dx_a}{dr}\right)^2 dr^2 + \sigma_{ij} dy^idy^j \nonumber \\ &=& \left[ h^2(r)  + s^2(r)\left(\frac{dx_a}{dr}\right)^2 \right] dr^2 + \sigma_{ij} dy^idy^j 
\end{eqnarray} The area of the surface is 
\begin{equation}
 A = \int_{r_0}^\infty V_{int} \left[ h^2(r)  + s^2(r)\left(\frac{dx_a}{dr}\right)^2 \right]^{1/2} dr. 
\end{equation} $V_{int}$ is the volume of the internal manifold.  We observe that the problem of finding the minimal area reduces to solving for the classical path of a point particle $x_a$.  Since the Lagrangian does not explicitly depend on the position of the particle, then the Euler Lagrange equations implies 
\begin{equation} \frac{d }{d r} \left[ \frac{\partial L}{\partial x_a'}\right] = 0 \;\;\; \mbox{where} \;\;\; L =  V_{int} \sqrt{ h^2(r)  + s^2(r)(x_a')^2 } 
\end{equation} where $'$ denotes derivative with respect to $r$.  The equation of motion for $x_a$ is given by 
\begin{equation} 
x_a'^2(s^2V_{int}^2 -c^2) = \frac{c^2 h^2}{s^2} \label{eq:3} 
\end{equation} where $c$ is an integration constant to be determined by the boundary condition at $r=r_0$.  However we note that the r.h.s. tends to infinity as $r\rightarrow r_0$.  This implies that the derivative of $x_a$ blows up at the horizon whenever $c\neq 0$.   The solution that minimize the action corresponds to then $c=0$, i.e.
\begin{equation}
x_a(r) = \mbox{constant}, 
\end{equation} and similarly for $x_b(r)$.  To match the boundary conditions at the conformal boundary, we find that the only allowed discontinuous surface for the space-times with zero volume end space is: 
\begin{equation}
\phi_1 = t-t_0, \;\;\;  \phi_2 = \begin{cases} x + l/2  \;\;\; \infty \leq r \leq r_0,  \\ r-r_0 \;\;\; a \leq x \leq b \\ x - l/2  \;\;\; r_0 \leq r \leq \infty\end{cases} . \label{eq:4} 
\end{equation}

\section{Geometric entropy from $AdS_4$:  $r(\theta)$ embedding}
\label{app:rtheta}

We start by proving that $\Delta A$ changes sign.  This quantity is given by,
\begin{equation}
\Delta A = 2 \beta L \left[ \int_{y_*}^{y_\infty} \left(\frac{y \sqrt{f(y)}}{\sqrt{y^2 f(y) -y_*^2 f(y_*)}} -1\right) dy - (y_* -y_0) \right].
\end{equation} First consider the case when $y_*$ is large; then $f(y)$ can be substitute for its asymptotic value of $y^2$.  We thus obtain
\begin{eqnarray}
\Delta A &=& 2 \beta L y_* \left[ \int_{1}^{x_\infty} \left(\frac{x^2}{\sqrt{x^4 -1}} -1\right) dx - 1 \right] \\
&=& 2 \beta L y_* \left[ .4009... - 1 \right]<0.
\end{eqnarray} So in the large $y_*$ limit, the difference in area is negative.

Now consider the $y_* \to y_0$ limit.  It is clear that
\begin{equation}
\Delta A (y_* =y_0) =0
\end{equation}
since $f(y_0)=0$.  Now we want to show that for $y_* = y_0 + \epsilon$, $\Delta A>0$.  For this we can deduce that the difference in area changes sign since it is positive for $y_*$ near $y_0$ and negative for large $y_*$.  We show this by proving that the derivative of $\Delta A$ with respect to $y_*$ is positive at $y_*$ and near $y_0$.  To facilitate taking derivatives, we rewrite the above expressions in terms of the coordinate $x=\frac{y_*}{y}$:
\begin{eqnarray}
f(y) &=& \frac{1}{y_* x^2} (y_*^3+ y_*x^2-x^3(y_0^3 + y_0)) \\
y^2 f(y) &=& \frac{y_*}{x^4} (y_*^3+ y_*x^2-x^3(y_0^3 + y_0)) \\
y^2 f(y) - y_*^2 f(y_*) &=& \frac{y_*}{x^4}(1-x) (x^3(y_*^3+y_* -y_0^3-y_0)+x^2 (y_*^3 +y_*)+y_*^3(1+x)).
\end{eqnarray}
In this coordinate, $\Delta A$ becomes,
\begin{eqnarray}
\Delta A &=& \mbox{\small $2 \beta L y_* \int_{0}^{1}\left(\frac{\sqrt{y_*^3+y_* x^2-\left(y_0^3+y_0\right) x^3}}{x^2\sqrt{1-x} \sqrt{\left(y_*^3+y_*-y_0^3-y_0\right) x^3+\left(y_*^3+y_*\right) x^2+(x+1)
   y_*^3}}-\frac{1}{x^2}\right)dx$} \nonumber \\  &-&2 \beta L (y_*-y_0). \label{da}
\end{eqnarray}
Taking derivative of $\Delta A$ with respect to $y_*$ at fixed $y_0$
\begin{eqnarray}
\frac{d \Delta A }{dy_*} &=&\mbox{\small $\beta L \int_{0}^1 \frac{x^2 y_* \left(-2 (x+1) y_*^3+3 \left(x^2+x+1\right) \left(y_0^3+y_0\right) y_*^2+x^2  \left(y_0^3+y_0\right)\right)dx}{\sqrt{1-x} \sqrt{-\left(y_0^3+y_0\right) x^3+y_*
   x^2+y_*^3} \left[\left(y_*^3+y_*-y_0^3-y_0\right) x^3+\left(y_*^3+y_*\right) x^2+(x+1)
   y_*^3\right]^{3/2}}$} \nonumber \\
&+& \frac{\Delta A}{y_*} - 2 \beta L \frac{y_0}{y_*}.
\end{eqnarray}
First, we observe that when $y_*=y_0$,
\begin{eqnarray}
\frac{d \Delta A }{dy_*} &=& \mbox{ \small $\beta L \int_{0}^1 \frac{x^2 y_0^2 (1+ 3y_0^2)(x^2+y_0^2(1+x+x^2))dx}{\sqrt{1-x}\sqrt{(1-x)(y_0x^2 + y_0^3(1+x+x^2))}[y_0x^2+y_0^3(1+x+x^2)]^{3/2}}$} - 2\beta L \nonumber\\
&=& \beta L \int_0^1 \frac{(1+3y_0^2)x^2}{(1-x)(x^2 + y_0^2(1+x+x^2))} - 2 \beta L.
\end{eqnarray}
We observe that the first term is positive and logarithmically divergent due to the $\frac{1}{1-x}$ factor in the integrand. This divergence comes from the fact that the term $ \sqrt{-\left(y_0^3+y_0\right) x^3+y_*x^2+y_*^3}$ in the denominator of the integrand in equation \ref{da} vanishes when $y_* =y_0$ and $x=1$.  Furthermore, one should note that all the other terms in the integrand are regular within the integration range except for the $\sqrt{1-x}$ term; in fact the integrand is zero at $x=0$.  Thus when $y_*$ is close to $y_0$, most of the integration will be coming from $x$ near 1 region.  To illustrate this picture better, we rewrite the integral as
\begin{equation}
\int_0^1 \frac{h(y_*,y_0,x) dx}{\sqrt{1-x}\sqrt{\frac{y_*}{y_0}-x}}
\end{equation} where $h(y_*,y_0,x)$ is regular for $x \in [0,1]$.  Notice that we have factored the term that leads to the divergence at $y_*=y_0$ since
\begin{equation}
-\left(y_0^3+y_0\right) x^3+y_*x^2+y_*^3 = (\frac{y_*}{y_0}-x)(y_0^3(\frac{y_*^2}{y_0^2} + \frac{y_*}{y_0}x + x^2)+y_0x^2).
\end{equation}
Written this way, it is clear that when $y_*$ is near $y_0$, most of the contribution of the integral comes from $x \approx 1$.  Thus for $y_*$ close to $y_0$, we can write
\begin{equation}
\frac{d \Delta A }{dy_*} = - \beta L h(1,y_0,y_0)\ln\left(\frac{y_*}{y_0}-1 \right) + \frac{\Delta A}{y_*}-2\beta L >0.
\end{equation} since $\frac{\Delta A}{y_*}$ is finite.

\section{A limiting value of $l(y_*)$: $r(\tau)$ embedding in $AdS_4$}

\label{sec:limitlaty0}
In this note we demonstrate that
\begin{equation}
l(y_* \to y_0) = \frac{1}{2}
\end{equation} for $r(\tau)$ in AdS$_4$.

The main quantity is given as
\begin{equation}
l(y_*) = \frac{2}{\tilde{\beta}}  \int_{y_*}^{y_\infty} \frac{y_* \sqrt{f(y_*)}}{f(y) \sqrt{y^2 f(y) - y_*^2 f(y_*)}} dy.
\end{equation}
we first transform to a more convenient coordinate $x = \frac{y_*}{y}$ where the various functions becomes,
\begin{eqnarray}
f(y) &=& \frac{1}{y_* x^2} (y_*^3+ y_*x^2-x^3(y_0^3 + y_0)) \\
y^2 f(y) &=& \frac{y_*}{x^4} (y_*^3+ y_*x^2-x^3(y_0^3 + y_0)) \\
y^2 f(y) - y_*^2 f(y_*) &=& \frac{y_*}{x^4}(1-x) (x^3(y_*^3+y_* -y_0^3-y_0)+x^2 (y_*^3 +y_*)+y_*^3(1+x)).
\end{eqnarray}  From these equations, we obtain

\begin{equation}
l(y_*) = \mbox{ \small $\frac{2 y_*^2\sqrt{y_*^3+y_* -y_0^3-y_0}}{\tilde{\beta}} \int_0^1 \frac{x^2 dx}{\sqrt{1-x}\sqrt{(x^3(y_*^3+y_* -y_0^3-y_0)+x^2 (y_*^3 +y_*)+y_*^3(1+x))}(y_*^3+ y_*x^2-x^3(y_0^3 + y_0))} $}.
\end{equation}

This function  has several characteristics that complicate the analysis near $y_0$.  These are
\begin{enumerate}
	\item The pre-factor of the integral vanishes when $y_*=y_0$.
	\item The denominator of the integrand have a zero when $x=1$.  Since this zero comes under a square, it is not a serious problem.
	\item When $y_* = y_0$, the integrand is unbounded in the $x=1$ limit.  From numerical analysis, we know that as $y_*$ approaches $y_0$, this unboundedness is such that the integral diverges the right way to cancel the zero in the pre-factor.
\end{enumerate}

The key to obtaining an analytic understanding of $l$ in the $y_* \to y_0$ limit, is to have a way of making case $3$ precise.  We can do this by factoring out the terms in the integral that are responsible for these pathologies.  We have

\begin{eqnarray}
y_*^3+y_* -y_0^3-y_0 &=& (y_* -y_0) (y_*^2 +y_*y_0 +y_0^2 +1)\\
y_*^3+y_*x^2 -x^3(y_0^3+y_0) &=& (y_* -y_0 x) (y_*^2 +x y_*y_0 +x^2 y_0^2 +x^2)
\end{eqnarray}

Now we can write $l(y_*)$ as
\begin{equation}
l(y_*) = \frac{2}{\tilde{\beta}}\int_0^1 \frac{\eta(x,y_*,y_0)}{\sqrt{1-x}}\frac{\sqrt{y_*-y_0}}{y_*-y_0 x} dx
\end{equation} where
\begin{equation}
\eta(x,y_*,y_0) = \frac{x^2 y_*^2 \sqrt{y_0^2+y_* y_0+y_*^2+1}}{\left(y_0^2 x^2+x^2+y_0 y_* x+y_*^2\right) \sqrt{\left(-y_0^3-y_0+y_*^3+y_*\right) x^3+\left(y_*^3+y_*\right) x^2+(x+1)
   y_*^3}}. \nonumber
\end{equation}

Now we have isolated all the pathologies and introduced the function $\eta$.  This function is everywhere regular.  This allows us to take limits in an organized way.  Let
\begin{equation}
\epsilon = \frac{y_* -y_0}{y_0}.
\end{equation}
Now we perform a set of coordinate transformations on the integral equation $l(y_*)$
\begin{eqnarray}
l &=& \frac{2}{\tilde{\beta}}\int_0^1 \frac{\eta(x,y_*,y_0)}{\sqrt{1-x}}\frac{\sqrt{y_*-y_0}}{y_*-y_0 x} dx = \frac{2\sqrt{\epsilon y_0}}{\tilde{\beta}}\int_0^1 \frac{\eta(x,y_*,y_0)}{\sqrt{1-x}}\frac{1}{y_*-y_0 x} dx \\
&=& \frac{2}{\tilde{\beta}}\sqrt{\frac{\epsilon}{ y_0}}\int_0^1 \frac{\eta(1-z,y_*,y_0)}{\sqrt{z}}\frac{1}{\epsilon +z} dz \;\;\; \mbox{where} \;\;\; x= 1-z
\end{eqnarray}
We now consider a coordinate change $z=u^2$ and the integral becomes

\be
l=\frac{4\sqrt{\epsilon}}{\tilde{\beta} \sqrt{y_0}} \int_0^1 \frac{\eta(1- u^2,y_*,y_0)}{\epsilon  + u^2} du
\ee
Using that
\be
\int\limits_0^1\frac{du}{u^2+\epsilon}=\frac{1}{\sqrt{\epsilon}}\arctan\left(\frac{1}{\sqrt{\epsilon}}\right),
\ee
we are ready to take the $\epsilon\to 0$ limit in the expression for $l$.
Now we let $y_*$ go to $y_0$ and $\epsilon$ go to 0.  We then obtain
\begin{equation}
l(y_* \to y_0) = \frac{4\eta(1,y_0,y_0)}{\tilde{\beta} \sqrt{y_0}} \frac{\pi}{2}.
\end{equation}
Now we have
\begin{eqnarray}
\eta(1,y_0,y_0) &=& \frac{y_0^2 \sqrt{1 + 3 y_0^2}}{(1 + 3 y_0^2)\sqrt{y_0 + 3 y_0^3}} \\
\tilde{\beta} &=& \frac{4 \pi y_0}{1 + 3 y_0^2}.
\end{eqnarray}
This allows us to conclude that

\begin{equation}
l(y_* \to y_0) = \frac{1}{2}.
\end{equation}

This sort of analytical result is very useful because it is used as a calibration for the numerics that we performed. In can clearly be seen from figure \ref{fig:ads4tl} that at the minimal value of $y_*$, which is $y_0$, the normalization of $l$ is precisely as analytically obtained above.

\end{document}